\documentclass[]{article}
\usepackage{amssymb,latexsym}
\usepackage{amsmath}
\usepackage{amsthm}
\usepackage{graphicx}
\usepackage{epsfig}
\usepackage{subfig}
\usepackage{fancyhdr}
\usepackage{tensor}
\usepackage{multicol}
\usepackage{cite}
\usepackage[left=3.5cm,top=3cm,bottom=3cm,right=3.5cm,columnsep=.6cm]{geometry}

\newcommand{\mc}{\mathcal}

\newcommand{\bs}{\begin{split}}
\newcommand{\es}{\end{split}}

\newcommand{\eqr}{\eqref}
\newcommand{\g}{g\indices}
\newcommand{\Q}{\mathcal{Q}}
\newcommand{\A}{\mathcal{A}}
\newcommand{\dr}{\ell}

\theoremstyle{plain}
\theoremstyle{definition}
\newtheorem{constraints}{Constraints}%[section]%[chapter]
\newtheorem{assertion}{Assertion}%[section]%[chapter]

\numberwithin{equation}{section}

\usepackage[colorlinks=true,linkcolor=blue,citecolor=blue,urlcolor=blue]{hyperref}
\begin{document}
%----------------------------------------------------------------------------------------------------------------------
\title{\textbf{Near-Extremal Kerr $AdS_2\times S^2$ Solution and Black-Hole/Near-Horizion-CFT Duality}}
\author{Ananda Guneratne$^{a}$\footnote{\href{mailto:guneratn@grinnell.edu}{guneratn@grinnell.edu}}, Leo Rodriguez$^{a}$\footnote{\href{mailto:rodrigul@grinnell.edu}{rodrigul@grinnell.edu}}, Sujeev Wickramasekara$^{a}$\footnote{\href{mailto:wickrama@grinnell.edu}{wickrama@grinnell.edu}}\\and Tuna Yildirim$^{b}$\footnote{\href{mailto:tuna-yildirim@uiowa.edu}{tuna-yildirim@uiowa.edu}}\vspace{.5cm}\\ \textit{$^{a}$Department of Physics}\\ \textit{Grinnell College}\\\textit{Grinnell, IA 50112}\vspace{.5cm}\\ \textit{$^{b}$Department of Physics and Astronomy}\\ \textit{The University of Iowa}\\\textit{Iowa City, IA 52242}}
\maketitle
%\date{\today}
\begin{abstract}
\noindent\rule{\linewidth}{0.5mm}
\noindent We study the thermodynamics of the near horizon of near extremal Kerr geometry ($near-NHEK$) within an $AdS_2/CFT_1$ correspondence.  We do this by shifting the horizon by a general finite mass, which does not alter the geometry and the resulting solution is still diffeomorphic to $NHEK$, however it allows for a Robertson Wilczek two dimensional Kaluza-Klein reduction and the introduction of a finite regulator on the $AdS_2$ boundary. The resulting asymptotic symmetry group of the two dimensional Kaluza-Klein reduction leads to a non-vanishing quantum conformal field theory ($CFT$) on the respective $AdS_2$ boundary. The $s$-wave contribution of the energy-momentum-tensor of the $CFT$, together with the asymptotic symmetries, generate a Virasoro algebra with calculable center and non-vanishing lowest Virasoro eigen-mode. The central charge and lowest eigen-mode reproduce the $near-NHEK$ Bekenstein-Hawking entropy via the statistical Cardy Formula and our derived central charge agrees with the standard Kerr/$CFT$ Correspondence. We also compute the Hawking temperature of the shifted $near-NHEK$ by analyzing quantum holomorphic fluxes of the Robinson and Wilczek two dimensional analogue fields.\\
\rule{\linewidth}{0.5mm}\vspace{.5cm}\vspace{.5cm}
\end{abstract}
%\begin{center}
%\noindent\line(1,0){150}
%\end{center}
\newpage
\thispagestyle{plain}
%----------------------------------------------------------------------------------------------------------------------
%\lhead{\rightmark}
%\chead{HW 10}
%\rhead{\thepage}
%\lfoot{A. Guneratne, L. Rodriguez, S. Wickramasekara, T. Yildirim}
%\cfoot{}
%\rfoot{\thepage}%\noindent We will take the speed of light $c=1$ through out this problem set.
%----------------------------------------------------------------------------------------------------------------------
%\begin{center}
%\rule{\linewidth}{0.5mm}
%\noindent\line(1,0){300}
%\end{center}
\tableofcontents
%\todototoc
%\listoftodos[List of Questions]
\begin{center}
\rule{\linewidth}{0.5mm}
%\noindent\line(1,0){300}
\end{center}
%\newpage
%------------------------------------------------------------------------------
\section{Introduction}\label{sec:intro}
%------------------------------------------------------------------------------
Black hole thermodynamic quantities \cite{hawk2,hawk3,beken},
\begin{align}
\label{eq:htaen}
\begin{cases}
T_H=\frac{\hbar\kappa}{2\pi}&\text{Hawking Temperature}\\
S_{BH}=\frac{A}{4\hbar G}&\text{Bekenstein-Hawking Entropy}
\end{cases},
\end{align}
provided an ample testing bed for most current competing theories of quantum gravity. It is widely believed that any viable ultraviolet completion of general relativity should reproduce some variant of \eqr{eq:htaen}, perhaps modulo some real finite parameter that would need to be fixed by experiment. To date there is a plethora of different approaches for arriving at \eqr{eq:htaen}, with string theories and loop quantum gravity the predominant competitors and no clear consensus of which approach should be preferred over the other.
%--
\subsection{Black Hole Temperature and Effective Action}\label{sec:bht}
%--
Since Hawking's original analysis of the density of quantum states in terms of Bogolyubov coefficients, analysis via effective actions and their associated energy momentum tensors of the semiclassical matter fields has been explored in various settings for arriving at $T_H$\cite{mukwipf,balfab2,balfab,cadtr,qpz,Camblong:2004ec,Camblong:2004ye,Yuan:2011gq,LLRphd}. Of particular interest is the realization by Robinson and Wilczek (RW) that anomalous two dimensional chiral theories in the near horizon of black holes are rendered unitary by requiring the black hole to radiate at temperature $T_H$ \cite{robwill,isowill,msoda,gango,Jin,Jinwu,chen,chen2,pwu,nampark,setare,petro,rabin3,rabin,rabin2,rabin4,Banerjee:2008sn,Wu:2011im,Akhmedova:2008au,Zampeli:2012tv,srv}. This procedure requires a dimensional reduction yielding two dimensional analogues for various types of four dimensional black holes (RW2DA), beyond the basic Schwarzschild case, coupled to two dimensional matter fields.

More recently Rodriguez and Yildirim showed by analyzing quantum conformal matter in the background of certain RW2DA black holes that the resulting energy momentum tensor (EMT) is holomorphic in the horizon limit or at the asymptotic infinity boundary \cite{ry,Button:2010kg}. Furthermore, this resulting holomorphic EMT is dominated by one component equaling the four dimensional Hawking flux of temperature $T_H$ weighted by a central charge $c$. A plausible interpretation of this result is that in the near horizon regime a four or higher dimensional spacetime metric exhibits a Kaluza-Klein reduction into two dimensional fields $\g{^{(2)}_\mu_\nu}$, $\mc{A}_\mu$ and $\Phi$. In this interpretation $\mc{A}_\mu$ and $\Phi$ are of gravitational origin, yet mathematically behave as two dimensional $U(1)$ gauge and conformal scalar fields. Thus a quantum field theoretic study of $\Phi$ in two dimensions with respect to $\g{^{(2)}_\mu_\nu}$ and $\mc{A}_\mu$ will have quantum gravitational implications in the near horizon of four dimensional black holes. Depending on the asymptotic symmetry group of certain RW2DA, this may suggest a sort of $AdS_2/CFT_1$ relationship for computing four dimensional black hole temperature.
%\cite{Bardeen:1999px}NHEKciteation
%--
\subsection{Holographic Black Hole Entropy}\label{sec:bhent}
%--
The fact that quantum gravity may be dual to a $CFT$ \cite{Maldacena:1997re} has spawned a surge, led by Strominger \cite{strom2,kerrcft}, Carlip \cite{Carlip:2011ax,carlip,carlip3,carlip2} and Park \cite{Park:1999tj,Park:2001zn,kkp}, in applying $CFT$ techniques to compute Bekenstein-Hawking Entropy of various black holes \cite{ry,Button:2010kg,cadss,silva,bgk,Barnes:2009zn,Astefanesei:2009sh,Banados:2011sd,Majhi:2011ws,Majhi:2012tf}. The most notable such example is by far the Kerr/$CFT$ correspondence and its extensions \cite{kerrcft,SheikhJabbaria:2011gc,deBoer:2011zt,Yavartanoo2012410,springerlink:10.1140/epjc/s10052-012-1911-7,rasmussen:2010xd,rasmussen:2010sa,Chen:2010yu,Chen:2010bh,Li:2010ch,Castro:2010fd,Krishnan:2010pv,kerrcftstring,kerrcftsugra,kerrcftind,daCunha:2010jj,Wu:2009di,Huang:2010yg,Compere:2012jk}, where the general idea is that the asymptotic symmetry group (ASG), preserving certain metric boundary or fall off conditions, is generated by a Virasoro algebra with calculable central extension:
\begin{align}
\label{eq:vir}
\left[\Q_m,\Q_n\right]=(m-n)\Q_{m+n}+\frac{c}{12}m\left(m^2-1\right)\delta_{m+n,0}.
\end{align}
The Bekenstein-Hawking entropy is then obtained from Cardy's Formula \cite{cardy2,cardy1} in terms of $c$ and the proper normalized lowest eigen-mode via:
\begin{align}
\label{eq:cf}
S=2\pi\sqrt{\frac{c\cdot\Q_0}{6}}.
\end{align}
In the Kerr/$CFT$ case a thermal Cardy Formula is employed 
\begin{align}
\label{eq:tcf}
S_{BH}=\frac{\pi^2}{3}c_LT_L,
\end{align}
which depends on the left Frolov-Thorne vacuum temperature $T_L$ for generic Kerr geometry \cite{frolovthorne}. This is in part due to the vanishing surface gravity of the extremal Kerr geometry, which is usually employed in regulating the quantum charges of \eqr{eq:vir}, thus leading to a finite $\Q_0^{L,R}$. However, a finite zero mode may be inferred by the identification:
\begin{align}\label{eq:fttdef}
\frac{\partial S_{CFT}}{\partial \Q_0}=\frac{\partial S_{BH}}{\partial \Q_0}=\frac{1}{T}\Rightarrow \Q_0=\frac{\pi^2}{6}c T^2.
\end{align}
Requiring that the ASG contains a proper $SL(2,\mathbb{R})$ subgroup in the above equation, yields the general value $T=\frac{1}{2\pi}$, which is the case for Kerr/$CFT$, where $T_L=\frac{1}{2\pi}$, $T_R=0$, and $c_L=c_R=c=12J$ and the extremal Bekenstein-Hawking entropy is recovered via the thermal Cardy Formula:
\begin{align}
\label{eq:tcf2}
S_{BH}=\frac{\pi^2}{3}c(T_L+T_R)=2\pi J.
\end{align} 
From the definition \eqr{eq:fttdef} we see that $T$ in general should be unitless. An interesting choice is to employ the Hawking temperature scaled by the finite time regulator $1/\kappa$ giving the general result $T=\frac{1}{2\pi}$ and also extends smoothly to extremality\footnote{A similar identification can be found in \cite{Carlip:2011ax,ChangYoung:2012kd}.}. 

The general Kerr entropy may be obtained by analyzing the Frolov-Thorne vacuum in the near extremal case, where 
\begin{align}\label{eq:nefttemp}
T_L=\frac{(GM)^2}{2\pi J}~\text{and}~T_R=\frac{\sqrt{(GM)^4-(GJ)^2}}{2\pi J}.
\end{align}
Using these values and $c=12J$ in \eqr{eq:tcf2} yields the standard area law 
\begin{align}\label{eq:mal}
S_{BH}=2\pi\left(GM^2+\sqrt{G^2M^4-J^2}\right).
\end{align}
Yet, the above result requires combining quantities derived separately at extremality and near-extremality. It is also not obvious that the combination of temperatures in \eqr{eq:nefttemp}, $T=T_L+T_R$, yields the value $\frac{1}{2\pi}$ except in the extremal limit. However recasting the values of $c,~T_L,~T_R$ into more general variables we see that:
\begin{align}\label{eq:nexctlr}
c=\frac{3A}{2\pi G},~T_{L}=\frac{4(GM)^2}{A}~\text{and}~T_{R}=\frac{4\sqrt{(GM)^4-(GJ)^2}}{A}
\end{align}
substituting these values into \eqr{eq:tcf} yields $S_{BH}=\frac{A}{4G}$ and assuming they smoothly extend back to non-extremality we have:
\begin{align}
T=T_{L}+T_{R}=\frac{4(GM)^2}{A}+\frac{4\sqrt{(GM)^4-(GJ)^2}}{A}=\frac{1}{2\pi},
\end{align}
which would provide a more wholesome computation of the near-extremal Kerr black hole entropy. This is precisely the aim of this note, to construct a near horizon $CFT$ dual for the $near-NHEK$ spacetime and compute its corresponding entropy within a statistical Cardy formula \eqr{eq:cf}, without mixing result derived separately at extremality and near-extremality. We will do this within an $AdS_2/CFT_1$ correspondence by performing a RW two dimensional reduction of the $near-NHEK$ geometry in a specific finite mass gauge following similar constructions as found in \cite{Button:2010kg,Castro:2009jf}. 

In \cite{Button:2010kg} Button, Rodriguez, Whiting and Yildirim showed that the RW2DA of non-extremal Kerr-Newman-$AdS$ is asymptotically $AdS_2$, for a specific choice of metric and gauge field fall off conditions, with effective near horizon functional derived via the RW dimensional reduction procedure.
%\begin{align}\label{eq:nhlcft}
%\bs
%S=&\frac{(r_+^2+a^2)}{16\pi G\Xi}\int d^2x\sqrt{-g^{(2)}}\left\{-\Phi\square_{g^{(2)}}\Phi+2\Phi R^{(2)}\right\}\\
%&+\frac{3 e^2 (r_+^2+a^2)}{\pi G\Xi}\int d^2x\sqrt{-g^{(2)}}\left\{-B\square_{g^{(2)}}B+2B \left(\frac{\epsilon^{\mu\nu}}{\sqrt{-g^{(2)}}}\right)\partial_\mu A_\nu\right\}.
%\es
%\end{align} 
Evaluating the resulting effective EMT on the $AdS_2$ boundary and computing its response to a total symmetry transformation yields a central charge:
\begin{align}\label{eq:centerp1}
\frac{c}{24\pi}=\frac{r_+^2+a^2}{4\pi G\Xi}\Rightarrow c=\frac{3A}{2\pi G},
\end{align}
which is in agreement with the Kerr/$CFT$ correspondence within the appropriate limits, i.e.
\begin{align}
\lim_{\dr\to\infty,~Q\to0,~a\to GM}c=12J.
\end{align}
Computation of the asymptotic symmetry algebra yields the normalized zero mode
\begin{align}
\Q_0=\frac{A}{16\pi G},
\end{align}
which together with the central charge and \eqr{eq:cf} yields the Bekenstein-Hawking Entropy of the KNAdS black hole. It is interesting to note that the lowest Virasoro eigen-mode satisfies
\begin{align}
\mathcal{Q}_0=GM_{irr}^2
\end{align}
where $M_{irr}^2$ is the irreducible mass of the KN$AdS$ black hole and agrees with \eqr{eq:nexctlr}. This suggests a possible addition to the $AdS/CFT$ dictionary, that the eigen-value of the $CFT$'s Hamiltonian is proportional to the irreducible mass of its black hole dual.

An additional distinct example of reducing near horizon dynamics to two dimensional physics was introduced by Castro and Larsen \cite{Castro:2009jf} by recasting the $NHEK$ \cite{Bardeen:1999px} into two dimensional fields $\g{^{(2)}_\mu_\nu}$, $\psi$ and $\mc{A}_\mu$
\begin{align}
\label{eq:clnhek}
ds^2=\frac{1+\cos^2{\theta}}{2}\left[ds_{(2)}^2+e^{-2\psi}\dr^2d\theta^2\right]+e^{-2\psi}\dr^2\frac{2\sin^2{\theta}}{1+\cos^2{\theta}}\left(d\phi+\mc{A}\right)^2
\end{align}
and evaluating the Einstein-Hilbert action over \eqr{eq:clnhek} and integrating out angular degrees of freedom.
%of freedom yielding the near horizon $AdS_2$ action:
%\begin{align}
%\label{eq:clnhekact}
%S=\frac{\dr^2}{4G}\int d^2x\sqrt{-g^{(2)}}\left\{e^{-2\psi}R^{(2)}+\frac{1}{\dr^2}+2\nabla_\mu e^{-\psi}\nabla^\mu e^{-\psi}+\frac{\dr^2}{2}e^{-4\psi}\mc{F}^2\right\}.
%\end{align}
A careful study of the asymptotic boundary currents of the resulting functional, within a well defined variational principle, yields a one dimensional quadratic two form with transformtion Law:
\begin{align}\label{eq:clemt}
\delta_{\epsilon+\Lambda}T_{tt}=&T_{tt}\xi'(t)+\xi(t)T'_{tt}+\frac{12J}{12}\dr\xi'''(t)+\mc{O}\left(e^{-\rho/\dr}\right)
\end{align}
in Gauss normal coordinates. This closely resembles the transformation law for the energy momentum tensor of a conformal field with center $c=12J$, in agreement with the Kerr/$CFT$ correspondence \cite{kerrcft}. However, a computation of the asymptotic symmetry algebra within this work was not performed. 

%--
%\subsection{Goals and Outline}\label{sec:FYI}
%--
We will proceed in Section~\ref{sec:nensea}, by reviewing the relevant $near-NHEK$ geometry and introduce the finite mass gauge by solving the vacuum Einstein field equations for generic four dimensional Einstein-Hilbert Theory with a specific initial ansatz. In Section~\ref{sec:qftnhnek} we perform the RW dimensional reduction to the geometry of interest and derive a two dimensional near horizon Liouville-like $CFT$ in terms of the resulting RW2DA fields. Analysis of the asymptotic symmetries of this theory leads to black hole thermodynamics of the near-extremal Kerr throat. Next, in Section~\ref{sec:2da}, we derive a normalized $AdS_2$ effective action via off shell analysis of the Einstein-Hilbert action directly within the finite mass gauge. The resulting boundary counter term contribution to the effective action yields a precise one dimensional quadratic two form with calculable central charge, reproducing the results of the previous section. Finally in Section~\ref{sec:concom} we close with a discussion of our results and hint towards future work.
%------------------------------------------------------------------------------
\section{Geometry}\label{sec:nensea}
%------------------------------------------------------------------------------
We are interested in studying four dimensional vacuum solutions which in the near horizon have the form:
\begin{align}
\label{eq:knadsnhm}
ds^2=K_1\left(\theta\right)\g{^{(2)}_\mu_\nu}dx^\mu dx^\nu+K_2\left(\theta\right)e^{-2\varphi}d\theta^2+K_3\left(\theta\right)e^{-2\varphi}\left[d\phi+\mathcal{A}\right]^2,
\end{align} 
The above two dimensional field splitting provides a robust platform for constructing $CFT$ duals for relevant classical spacetimes with non vanishing surface gravity (near-extremal), by analyzing the ASG of the Kaluza-Klein fields $\g{^{(2)}_\mu_\nu}$, $\mc{A}$ and $\varphi$. The $NHEK$ is a four dimensional vacuum solution, which is derived by taking the extremal near horizon limit:
\begin{align}
\label{eq:nhekcoor}
r=GM+\lambda U,~t'=\frac{t}{\lambda},~\phi'=\phi+\frac{t}{2GM\lambda},~\lambda\to\infty
\end{align}
of the generic Kerr metric:
\begin{align}
\label{eq:kerrm}
\bs
ds^2_{Kerr}=&-\frac{\Sigma\Delta}{\left(r^2+a^2\right)^2-\Delta a^2\sin^2\theta}dt'^2+\Sigma\left[\frac{dr^2}{\Delta}+d\theta^2\right]\\
&+\frac{\left(\left(r^2+a^2\right)^2-\Delta a^2\sin^2\theta\right)\sin^2\theta}{\Sigma}\left[d\phi'+\frac{2rGMa}{\left(r^2+a^2\right)^2-\Delta a^2\sin^2\theta}dt'\right]^2,
\es
\end{align}
where
\begin{align}
\label{eq:kerrdef}
\bs
\Sigma=&r^2+a^2\cos^2\theta,\\
\Delta=&\left(r-r_+\right)\left(r-r_-\right),\\
r_\pm=&GM\pm\sqrt{(GM)^2-a^2},\\
a=&\frac{J}{M},
\es
\end{align}
which yields:
\begin{align}
\label{eq:nhek}
ds^2_{NHEK}=\frac{1+\cos^2{\theta}}{2}\left[-\frac{U^2}{\dr^2}dt^2+\frac{\dr^2}{U^2}dU^2+\dr^2d\theta^2\right]+\dr^2\frac{2\sin^2{\theta}}{1+\cos^2{\theta}}\left(d\phi+\frac{U}{\dr^2}dt\right)^2.
\end{align}
The above metric is of the form \eqr{eq:knadsnhm}, and may be tuned to near-extremality via the finite temperature gauge yielding:
\begin{align}
\label{eq:nnhek}
ds^2_{near-NHEK}=\frac{1+\cos^2{\theta}}{2}\left[-\frac{U^2-\epsilon^2}{\dr^2}dt^2+\frac{\dr^2}{U^2-\epsilon^2}dU^2+\dr^2d\theta^2\right]+\dr^2\frac{2\sin^2{\theta}}{1+\cos^2{\theta}}\left(d\phi+\frac{U}{\dr^2}dt\right)^2,
\end{align}
where $\epsilon=\frac{1}{2\lambda}\left(r_+-r_-\right)$ is a finite excitation above extremality and $\dr^2=2G^2M^2$. 
%--
\subsection{Finite Mass Gauge}\label{sec:nhgeo}
%--
To aid in our $CFT$ dual construction we want to endow the $near-NHEK$ geometry with a finite $ADM$ mass and angular momentum parameter $M$ and $a$. We will do this by solving the Einstein Field Equations directly for specific boundary conditions, as opposed to implementing a parameter tuning process in the transformations of \eqr{eq:nhekcoor}. Starting with \eqr{eq:knadsnhm} as our ansatz, we impose the following symmetry conditions:
\begin{constraints}[Black Hole Symmetries]~
\begin{itemize}
\item The horizon is topologically $S^2$, i.e.
\begin{align}
\label{eq:hs2c}
K_2K_3=\sin^2\theta~\text{and}~K_1=K_2.
\end{align}
\item The metric is maximally isometric in $t$ and $\phi$, which implies
\begin{align}
\label{eq:misotphi}
\g{^{(2)}_{\mu\nu}}=\g{^{(2)}_{\mu\nu}}(r),~\varphi=\varphi(r)~\text{and}~\A=\A_\mu(r)dx^\mu.
\end{align}
\item The metric is axially symmetric in four dimensions and spherically symmetric in two:
\begin{align}
\label{eq:axisphs}
ds^2=\g{^{(2)}_{\mu\nu}}dx^\mu dx^\nu=-f(r)dt^2+\frac{dr^2}{f(r)}~\text{and}~\A=\A_t(r)dt.
\end{align}
\end{itemize}
\end{constraints}
\noindent Implementing these conditions within the four dimensional vacuum Einstein field equations we obtain the relatively simple set of coupled differential equations:
\begin{align}
\label{eq:refeq}
\begin{cases}
\frac{\varphi '(r) K_2'(\theta )}{K_2(\theta )}=0\\
~\\
K_2(\theta ) \left(-2 e^{-2 \varphi(r)} f'(r) \varphi'(r)+f(r)\left(4e^{-2 \varphi(r)}\varphi'(r)^2-2e^{-2\varphi(r)}\varphi''(r)\right)-2\right)+\\
K_2''(\theta )-3\cot(\theta)K_2'(\theta)=0\\
~\\
\sin (\theta ) \left(e^{-4 \varphi (r)} \sin ^2(\theta ) \A_t'(r)^2+K_2'(\theta)^2\right)+\sin (\theta ) K_2(\theta )^2 \left(-2 e^{-2 \varphi (r)} f'(r) \varphi'(r)\right.+\\
\left.f(r)\left(4 e^{-2 \varphi (r)} \varphi '(r)^2-2 e^{-2 \varphi (r)} \varphi''(r)\right)-2\right)-K_2(\theta ) \left(\sin (\theta ) K_2''(\theta )+\cos (\theta) K_2'(\theta )\right)=0\\
~\\
e^{-6 \varphi (a)} \sin ^2(\theta ) \A_t'(r)^2-e^{-4 \varphi (r)} K_2(\theta )^2 f''(r)-e^{-2
   \varphi (r)} \left(K_2(\theta )^2 \left(2 f(r) \left(4 e^{-2 \varphi (r)} \varphi '(r)^2-\right.\right.\right.\\
\left.\left.\left.2 e^{-2
   \varphi (r)} \varphi ''(r)\right)-2 e^{-2 \varphi (r)} f'(r) \varphi '(r)\right)-K_2'(\theta
   )^2+K_2(\theta ) \left(K_2''(\theta )+\cot (\theta ) K_2'(\theta
   )\right)\right)+\\
4 f(r) e^{-4 \varphi (r)} K_2(\theta )^2 \varphi '(r)^2=0
\end{cases}
\end{align}
where prime denotes derivation with respect to the argument of the given function. A general solution to the above field equations is given by:
\begin{align}
\label{eq:gbhs}
\begin{cases}
\varphi(r)=&\varphi\\
K_2(\theta)=&\frac{C_1 (1-\cos (\theta ))^{3/2} \sqrt{\cos ^2(\theta )-1}}{\sqrt{\cos (\theta )+1}}-\frac{C_2 \sqrt{1-\cos (\theta )} \cos (\theta )
   \sqrt{\cos ^2(\theta )-1}}{(\cos (\theta )-1) \sqrt{\cos (\theta )+1}}\\
\A_t(r)=&\frac{r \sqrt{C_2^2-4 C_1 C_2}}{e^{-2\varphi}}+C_3\\
f(r)=&\frac{r^2}{e^{-2\varphi}}+r C_5+C_4
\end{cases}
\end{align}
where $C_i$'s are integration constants and the full line element is given in \eqr{eq:knadsnhm} and reads:
\begin{align}
\label{eq:knadsnhmgs}
\bs
ds^2=&\left(\frac{C_1 (1-\cos (\theta ))^{3/2} \sqrt{\cos ^2(\theta )-1}}{\sqrt{\cos (\theta )+1}}-\frac{C_2 \sqrt{1-\cos (\theta )} \cos (\theta )\sqrt{\cos ^2(\theta )-1}}{(\cos (\theta )-1) \sqrt{\cos (\theta )+1}}\right)\cdot\\
&\left[-\left(\frac{r^2}{e^{-2\varphi}}+r C_5+C_4\right)dt^2+\frac{1}{\frac{r^2}{e^{-2\varphi}}+r C_5+C_4}dr^2+e^{-2\varphi}d\theta^2\right]+\\
&\frac{\sin^2(\theta)}{\frac{C_1 (1-\cos (\theta ))^{3/2} \sqrt{\cos ^2(\theta )-1}}{\sqrt{\cos (\theta )+1}}-\frac{C_2 \sqrt{1-\cos (\theta )} \cos (\theta )
   \sqrt{\cos ^2(\theta )-1}}{(\cos (\theta )-1) \sqrt{\cos (\theta )+1}}}\cdot\\
&e^{-2\varphi}\left[d\phi+\left(\frac{r \sqrt{C_2^2-4 C_1 C_2}}{e^{-2\varphi}}+C_3\right)dt\right]^2
\es
\end{align}

Next, we impose the final set of boundary conditions to fix the integration constants in the solution above:
\begin{constraints}\label{con:kerr}[Finite Mass Gauge]~
\begin{itemize}
\item The inner and outer horizons are located at:
\begin{align}
\label{eq:csh}
f(r_\pm)=0~\text{and}~r_\pm=GM\pm\sqrt{(GM)^2-a^2}.
\end{align}
\item The ADM mass is non zero and equal to the parameter $M$:
\item The total angular momentum is given by $J=\frac{a^2+r^2_+}{2G}$.
\end{itemize}
\end{constraints}
\noindent Applying these conditions yields the final solution to \eqr{eq:refeq}
\begin{align}
\label{eq:kbhs}
\begin{cases}
e^{-2\varphi}=&r_+^2+a^2\\
K_2(\theta)=&\frac{1+\cos^2\theta}{2}\\
\A_t(r)=&\frac{r-2GM}{r_+^2+a^2}\\
f(r)=&\frac{r^2-2rGM+a^2}{r_+^2+a^2}
\end{cases},
\end{align}
with line element 
\begin{align}
\label{eq:knadsnhmkerr}
\bs
ds^2=&\frac{1+\cos^2\theta}{2}\left[-\frac{r^2-2rGM+a^2}{r_+^2+a^2}dt^2+\frac{r_+^2+a^2}{r^2-2rGM+a^2}dr^2+\left(r_+^2+a^2\right)d\theta^2\right]+\\
&\frac{2\sin^2\theta}{1+\cos^2\theta}\left(r_+^2+a^2\right)\left[d\phi+\left(\frac{r-2GM}{r_+^2+a^2}\right)dt\right]^2.
\es
\end{align}
The above line element clearly exhibits global $AdS_2\times S^2$ topology and is diffeomorphic to \eqr{eq:nhek} and \eqr{eq:nnhek} as we will demonstrate shortly. However, the above line element is written in Boyer-Lindquist type coordinates and exhibits two physical parameters $M$ and $a$, which will be useful for tuning purposes in our CFT construction.
%--
\subsection{Surface Gravity}\label{sec:NHnEKsg}
%--
To compute the surface gravity of the gauged $near-NHEK$ ($gnNHEK$) we will exploit the fact that it is maximally isometric with respect to the coordinates $t$ and $\phi$ and define the general Killing vector:
\begin{align}
\label{eq:gkvec}
\xi=\xi_{(t)}+\Omega_H\xi_{(\phi)},
\end{align}
where $\Omega_H$ is the minimum of the function 
\begin{align}
\label{}
\frac{d\phi}{dt}=-\frac{g_{t\phi}}{g_{\phi\phi}}\pm\sqrt{\left(\frac{g_{t\phi}}{g_{\phi\phi}}\right)^2-\frac{g_{tt}}{g_{\phi\phi}}}
\end{align}
evaluated on the horizon $r_+$. Given this above Killing vector and making use of the geodetic equation to rearrange Frobenius' theorem
\begin{align}
\label{eq:fbt}
\nabla_{[\alpha}\xi_{\mu}\xi_{\nu]}=0
\end{align}
for hypersurface orthogonal congruences of null generators, we obtain:
\begin{align}
\label{eq:fbtsg}
\kappa^2=-\frac12\left.\nabla^\mu\xi^\nu\nabla_\mu\xi_\nu\right\vert_{r_+}.
\end{align}
Evaluating this over the connection of \eqr{eq:knadsnhmkerr} yields:
\begin{align}
\label{eq:sgnhnek}
\kappa_{gnNHEK}=\frac12f'(r_+)=\frac{r_+-GM}{r_+^2+a^2},
\end{align}
which leaves us with a finite non zero value, but may be tuned to zero for the case when $a\to GM$ i.e. when we approach extremality. We will also note the horizon are of the $gnNHEK$ black hole, which comes from evaluating the integral
\begin{align}
A=\int d^2x\sqrt{\gamma}=4\pi\left(r_+^2+a^2\right),
\end{align}
where $\gamma_{ab}$ are the metric degrees of freedom left over in \eqr{eq:knadsnhmkerr} after setting $dr=dt=0.$
%--
\subsection{$NHEK$-Map}\label{sec:NHEKl}
%--
As mentioned above, the solution \eqr{eq:knadsnhmkerr} is a $near-NHEK$ solution which can be mapped into \eqr{eq:nhek} via a similar coordinate relationship between $U$ and $r$ as in \cite{Bardeen:1999px}. Starting with \eqr{eq:knadsnhmkerr} and applying the coordinate redefinitions:
\begin{align}
\label{eq:nhekcoor24}
r=GM+\lambda U,~t=\frac{\tilde t}{\lambda},~\phi=\tilde \phi+\frac{\tilde t}{2GM\lambda},
\end{align}
and in stead of taking the limit as $\lambda\to0$ we set $a\to GM$ resulting in the $NHEK$ solution:
\begin{align}
\label{eq:nhekfus}
ds^2=\frac{1+\cos^2{\theta}}{2}\left[-\frac{U^2}{\dr^2}d\tilde t^2+\frac{\dr^2}{U^2}dU^2+\dr^2d\theta^2\right]+\dr^2\frac{2\sin^2{\theta}}{1+\cos^2{\theta}}\left(d\tilde \phi+\frac{U}{\dr^2}d\tilde t\right)^2.
\end{align}
In fact, as discussed in \cite{Amsel:2009ev,Maldacena:1998uz,Bredberg:2009pv}, there always exists a coordinate mapping between a $near-NHEK$ and a $NHEK$ type solution as evident in the form of their Kretschmann invariants:
\begin{align}
R_{\alpha\mu\beta\nu}R^{\alpha\mu\beta\nu}=\frac{1536 \sin ^2\theta (-52 \cos (2 \theta )+\cos (4 \theta )-45)}{(\cos (2 \theta )+3)^6 \dr^4}
\begin{cases}
\dr^2=2G^2M^2& NHEK\\
\dr^2=2G^2M^2& near-NHEK\\
\dr^2=r^2_++a^2& gnNHEK
\end{cases}.
\end{align} 
Notice, that for the choice $a=GM$ all invariants exhibit $\dr^2=2G^2M^2$, i.e. the extremal limit of \eqr{eq:knadsnhmkerr}.
%------------------------------------------------------------------------------
\section{Quantum Fields in $gnNHEK$ Spacetime}\label{sec:qftnhnek}
%------------------------------------------------------------------------------
We will now study the resulting near horizon matter theory via the RW dimensional reduction procedure. Our goal, for this section, will be to apply our previous techniques from \cite{ry,Button:2010kg} to study the resulting thermodynamics of the $gnNHEK$ within an $AdS_2/CFT_1$ formalism. 
%--
\subsection{RW Dimensional Reduction}\label{sec:RWdr}
%--
In our initial ansatz leading to the $gnNHEK$ solution we assumed a specific decomposition of our four dimensional spacetime into two dimensional black hole and matter fields. However, we have not shown that these fields are the correct RW2DA useful in a holographic study of the quantum spacetime in the near horizon regime.

 Let us consider a single free scalar field in the background of \eqr{eq:knadsnhmkerr} with functional:
\begin{align}
\label{eq:freescalar4}
\bs
S_{free}=&\frac12\int d^4x\sqrt{-g}g^{\mu\nu}\partial_{\mu}\varphi\partial_\nu\varphi\\
=&-\frac12\int d^4x\,\varphi\left[\partial_\mu\left(\sqrt{-g}g^{\mu\nu}\partial_\nu\right)\right]\varphi\\
=&-\frac12\int d^4x\,\varphi\left[\partial_t\left( -\sin{\theta}\left(a^2+r_+^2\right)\frac{r_+^2+a^2}{r^2-2rGM+a^2}\partial_t\right)+\right.\\
&\partial_r\left( \sin{\theta}\left(a^2+r_+^2\right)\frac{r^2-2rGM+a^2}{r_+^2+a^2}\partial_r\right)+\partial_\theta\left( \sin{\theta}\partial_\theta\right)+\\
&\partial_\phi\left( \left\{-\sin{\theta}\left(a^2+r_+^2\right)\left(\frac{r-2GM}{r_+^2+a^2}\right)^2\frac{r_+^2+a^2}{r^2-2rGM+a^2}+\frac{(\cos{(2 \theta )}+3)^2}{16\sin\theta}\right\}\partial_\phi\right)+\\
&\left.2\partial_t\left( \sin{\theta}\left(a^2+r_+^2\right)\frac{r-2GM}{r_+^2+a^2}\frac{r_+^2+a^2}{r^2-2rGM+a^2}\partial_\phi\right)\right]\varphi.
\es
\end{align}
The above functional is reduced to a two dimensional theory by expanding the four dimensional scalar field in terms of spherical harmonics
\begin{align}
\label{eq:sphdecom}
\varphi(t,r,\theta,\phi)=\sum_{lm}\varphi_{lm}(r,t)Y\indices{_l^m}(\theta,\phi),
\end{align}
where $\varphi_{lm}$ is a complex interacting two dimensional scalar field, and integrating out angular degrees of freedom. Transforming to tortoise coordinates defined as:
\begin{align}
\label{eq:ttcoor}
\frac{dr^{*}}{dr}=\frac{1}{f(r)}
\end{align}
and considering the region very close to $r_+$ we find the two dimensional action is much reduced since all interaction, mixing and potential terms ($\sim l(l+1)\ldots$) are weighted by a factor of $f(r(r*))\sim e^{2\kappa r^{*}}$, which vanishes exponentially fast as $r\to r_+$. This leaves us with an infinite collection of massless charged scalar fields in the very near horizon region, with $U(1)$ gauge charge $m$ and remnant functional:
\begin{align}
\label{eq:nhapw}
S=-\frac{r_+^2+a^2}{2}\int d^2x\;\varphi^{*}_{lm}\left[-\frac{1}{f(r)}\left(\partial_t-im\A_t\right)^2+\partial_rf(r)\partial_r\right]\varphi_{lm}.
\end{align}
Thus, we arrive at the RW2DA for the $ngNHEK$ solution given by:
\begin{align}
\label{eq:2drwamet}
\g{^{(2)}_\mu_\nu}=\left(\begin{array}{cc}-f(r) & 0 \\0 & \frac{1}{f(r)}\end{array}\right)
\end{align}
and $U(1)$ gauge field
\begin{align}
\label{eq:rw2dgf}
\A=\A_tdt.
\end{align}

Given the initial ansatz \eqr{eq:knadsnhm}, it is not surprising that the only relevant physical fields in the region $r\sim r_+$ are the RW2DAs above and makes the holographic statement that we may learn much about the quantum nature of spacetime in the near horizon regime via the semiclassical analysis of $\g{^{(2)}_\mu_\nu}$, $\A$ and $\varphi_{lm}$.
%--
\subsection{Effective Gravitational Action and Asymptotic Symmetries}\label{sec:asefa}
%--
We would like to interpret \eqr{eq:nhapw} as a useful action for gravity in the near horizon of classical four dimensional spacetime. This can be done by only considering the $s-wave$ contribution and making a field redefinition rendering the scalar field unitless \cite{Yale:2010tn,Chung:2010xy,Chung:2010xz,solodukhin:1998tc}. The $s-wave$ approximation is sensible in this scenario, since we will interpret $\varphi_{lm}$ as a component of the gravitational field and hence it should be real and unitless. Most of the interesting gravitational dynamics seem to be contained in this region or approximation \cite{strom1}\footnote{In \cite{Button:2010kg} it was shown that $\varphi_{lm}$ dies exponentially fast in time by analyzing the asymptotic behavior of its field equation. However we find the statement relating $\varphi_{lm}$ to a real gravitational field component, a stronger justification to neglect higher order terms in $l$ and $m$.}. Motivated by these arguments we make the field redefinition
\begin{align}\label{eq:sfrd}
\varphi_{00}=\sqrt{\frac{6}{G}}\psi,
\end{align} 
where $\psi$ is now unitless and the $\sqrt{6}$ was chosen to recover the Einstein coupling $\frac{1}{16\pi G}$ in the quantum gravitational effective action of \eqr{eq:nhapw} within the $s$-wave approximation.
Applying the field redefinition \eqr{eq:sfrd} to \eqr{eq:nhapw} yields:
\begin{align}\label{eq:tdtrw2}
S^{(2)}[\psi,g]=\frac{3(r_+^2+a^2)}{G}\int d^2x\sqrt{-g^{(2)}}\psi\left[D_{\mu}\left(\sqrt{-g^{(2)}}\g{_{(2)}^{\mu\nu}}D_{\nu}\right)\right]\psi,
\end{align}
where $D_\mu$ is the gauge covariant derivative. In addition to redressing the scalar field, our choice of field redefinition has also rendered the effective coupling unitless, hinting towards a finite quantum theory. The effective action of this quantum theory, which may be extracted via zeta-function regularization of the functional determinant in \eqr{eq:tdtrw2}, is given by the sum of two functionals \cite{isowill,Leutwyler:1984nd}:
\begin{align}\label{eq:nhpcft}
\Gamma=&\Gamma_{grav}+\Gamma_{U(1)},
\end{align}
where
\begin{align}
\bs
\Gamma_{grav}=&\frac{(r_+^2+a^2)}{16\pi G\Xi}\int d^2x\sqrt{-g^{(2)}}R^{(2)}\frac{1}{\square_{g^{(2)}}}R^{(2)}~\mbox{and}\\
\Gamma_{U(1)}=&\frac{3 e^2 (r_+^2+a^2)}{\pi G\Xi}\int \mc{F}\frac{1}{\square_{g^{(2)}}}\mc{F}.
\es
\end{align}
Next,  we introduce the auxiliary scalars $\Phi$ and $B$ satisfying:
\begin{align}\label{eq:afeqm}
\square_{g^{(2)}} \Phi=R~\mbox{and}~\square_{g^{(2)}} B=\epsilon^{\mu\nu}\partial_\mu \A_\nu,
\end{align}
which transforms the functional \eqr{eq:nhpcft} into a Liouville $CFT$ of the form:
\begin{align}\label{eq:nhlcft}
\bs
S_{NHCFT}=&\frac{(r_+^2+a^2)}{16\pi G\Xi}\int d^2x\sqrt{-g^{(2)}}\left\{-\Phi\square_{g^{(2)}}\Phi+2\Phi R^{(2)}\right\}\\
&+\frac{3 e^2 (r_+^2+a^2)}{\pi G\Xi}\int d^2x\sqrt{-g^{(2)}}\left\{-B\square_{g^{(2)}}B\right.\\
&+\left.2B \left(\frac{\epsilon^{\mu\nu}}{\sqrt{-g^{(2)}}}\right)\partial_\mu A_\nu\right\}
\es
\end{align}

Now, we turn our attention to computing the ASG of \eqr{eq:tdtrw2}. The behavior of the RW2DA fields at large $r$ is defined by
\begin{align}\label{eq:as2dwa}
\g{^{(0)}_\mu_\nu}=&
\left(
\begin{array}{cc}
-\frac{r^2}{\dr^2}+\frac{2 r G M}{\dr^2}-\frac{a^2}{\dr^2}+\mathcal{O}\left(\left(\frac{1}{r}\right)^3\right)& 0 \\
 0 & \frac{\dr^2}{r^2}+\mathcal{O}\left(\left(\frac{1}{r}\right)^3\right) 
\end{array}
\right),\\
\label{eq:asgf}
\mathcal{A}\indices{^{(0)}_t}=&\frac{r}{\dr^2}-\frac{2 G M}{\dr^2}+\mathcal{O}\left(\left(\frac{1}{r}\right)^3\right),
\end{align}
which yield an asymptotically $AdS_2$ configuration with Ricci Scalar, $R=-\frac{2}{l^2}+O\left(\left(\frac{1}{r}\right)^1\right)$, where $\dr^2=r_+^2+a^2$. In addition, we impose the following metric and gauge field fall-off conditions:
\begin{align}\label{eq:mbc}
\delta g_{\mu\nu}=
\left(
\begin{array}{cc}
    \mathcal{O}\left(\left(\frac{1}{r}\right)^3\right)&
   \mathcal{O}\left(\left(\frac{1}{r}\right)^0\right) \\
 \mathcal{O}\left(\left(\frac{1}{r}\right)^0\right) &
\mathcal{O}\left(r\right) 
\end{array}
\right)~\mbox{and}~\delta \mathcal{A}=\mathcal{O}\left(\left(\frac{1}{r}\right)^0\right),
\end{align}
which imply the following set of asymptotic metric preserving diffeomorphisms:
\begin{align}\label{eq:dpr}
\xi_n=\xi_1(r)\frac{e^{i n \kappa\left(t\pm r^*\right)}}{\kappa}\partial_t+\xi_2(r)\frac{e^{i n \kappa\left(t\pm r^*\right)}}{\kappa}\partial_r,
\end{align}
where $r^*$ is the tortoise coordinate defined by $dr^*=\frac{1}{f(r)}dr$,
\begin{align}
\xi_1=C r e^{i n \kappa  r^*},~\xi_2=\frac{i r C (r-G M)}{\kappa  n \left(r^2-2 r G M+a^2\right)},
\end{align}
$C$ is an arbitrary normalization constant and $\kappa$ is the surface gravity of the $gnNHEK$ black hole. Under this set of diffeomorphisms the gauge field transforms as:
\begin{align}
\delta_\xi \mathcal{A}_\mu=\left(\mathcal{O}\left(\left(\frac{1}{r}\right)^0\right),\mathcal{O}\left(\left(\frac{1}{r}\right)^1\right)\right)
\end{align}
and thus $\delta_{\xi}$ may be elevated to a total symmetry of the action, i.e.
\begin{align}
\delta_\xi\rightarrow\delta_{\xi+\Lambda},
\end{align}
in accordance with \eqr{eq:mbc}. Switching to light cone coordinates $x^\pm=t\pm r^*$,\footnote{Large $r$ behavior will be synonymous with large $x^+$ behavior.}
we see that the set $\xi_n^\pm$ is well behaved on the $r\rightarrow\infty$ boundary and form a centerless Witt or $Diff(S^1)$ subalgebra:
\begin{align}
i\left\{\mathbf{\xi}^\pm_m,\mathbf{\xi}^\pm_n\right\}=(m-n)\mathbf{\xi}^\pm_{m+n}.
\end{align}
%--
\subsection{Energy Momentum and The Virasoro algebra}\label{sec:thermo}
%--
We define the energy momentum tensor and $U(1)$ current of \eqr{eq:nhlcft} in their usual ways:
\begin{align}
\label{eq:emt}
\bs
\left\langle T_{\mu\nu}\right\rangle=&\frac{2}{\sqrt{-g^{(2)}}}\frac{\delta S_{NHCFT}}{\delta g\indices{^{(2)}^\mu^\nu}}\\
=&\frac{r_+^2+a^2}{8\pi G}\left\{\partial_\mu\Phi\partial_\nu\Phi-2\nabla_\mu\partial_\nu\Phi+g\indices{^{(2)}_\mu_\nu}\left[2R^{(2)}-\frac12\nabla_\alpha\Phi\nabla^\alpha\Phi\right]\right\}\\
&+\frac{6 e^2 (r_+^2+a^2)}{\pi G}\left\{\partial_\mu B\partial_\nu B-\frac12\g{_\mu_\nu}\partial_\alpha B\partial^\alpha B\right\}~\mbox{and}\\
\left\langle J^{\mu}\right\rangle=&\frac{1}{\sqrt{-g^{(2)}}}\frac{\delta S_{NHCFT}}{\delta \mathcal{A}_\mu}=\frac{6 e^2 (r_+^2+a^2)}{\pi G}\frac{1}{\sqrt{-g^{(2)}}}\epsilon^{\mu\nu}\partial_\nu B
\es
\end{align}
Next, solving the equation of motions for the auxiliary fields:
\begin{align}
\label{eq:eqmp}
\bs
\square_{g^{(2)}}\Phi=&R^{(2)}\\
\square_{g^{(2)}}B=&\epsilon^{\mu\nu}\partial_\mu \mathcal{A}_\nu
\es
\end{align}
using the metric \eqr{eq:2drwamet} and gauge field \eqr{eq:rw2dgf} and employeeing modified Unruh Vacuum boundary conditions (MUBC) \cite{unruh}
\begin{align}
\label{eq:ubc}
\begin{cases}
\left\langle T_{++}\right\rangle=\left\langle J_{+}\right\rangle=0&r\rightarrow\infty,~\dr\rightarrow\infty\\
\left\langle T_{--}\right\rangle=\left\langle J_{-}\right\rangle=0&r\rightarrow r_+
\end{cases},
\end{align}
we determine all relevant integration constants of \eqr{eq:emt} and \eqr{eq:eqmp}. For large $r$ and  to $\mathcal{O}(\frac{1}{\dr})^2$, which we will denote as the single limit $r\to\infty$ in the remainder of this section, the resulting energy momentum tensor is dominated by one holomorphic component, $\left\langle T_{--}\right\rangle$. We are interested in the response of the energy momentum tensor and the $U(1)$ current to a total symmetry $\delta_{\xi^-_n+\Lambda}$, which after expanding in terms of the boundary fields \eqr{eq:as2dwa} and \eqr{eq:asgf} we obtain:
\begin{align}
\begin{cases}
\delta_{\xi^-_n+\Lambda}\left\langle T_{--}\right\rangle=\xi^-_n\left\langle T_{--}\right\rangle'+2\left\langle T_{--}\right\rangle\left(\xi^-_n\right)'+\frac{r_+^2+a^2}{4\pi G}\left(\xi^-_n\right)'''+\mathcal{O}\left(\left(\frac{1}{r}\right)^3\right)\\
\delta_{\xi^-_n+\Lambda}\left\langle J_{-}\right\rangle=\mathcal{O}\left(\left(\frac{1}{r}\right)^3\right)
\end{cases}
\end{align}
From the above we see that $\left\langle T_{--}\right\rangle$ transforms asymptotically as the energy momentum tensor of a one dimensional $CFT$ with center:
\begin{align}\label{eq:center}
\frac{c}{24\pi}=\frac{r_+^2+a^2}{4\pi G}\Rightarrow c=\frac{3A}{2\pi G},
\end{align}
We should also note that the above central charge is in congruence with the 2-dimensional conformal/trace anomaly \cite{cft}:
\begin{align}
\label{eq:tra}
\left\langle T\indices{_\mu^\mu}\right\rangle=-\frac{c}{24\pi}R^{(2)}
\end{align}

Next, we define the quantum generators via the conserved charge:
\begin{align}
\label{eq:ccppb}
\Q_n=\lim_{r\rightarrow\infty}\int dx^-\left\langle T_{--}\right\rangle\mathbf{\xi}^-_n,
\end{align}
Whose algebraic structure is revealed by computing its response to a total symmetry, while compactifying the $x^-$ coordinate to a circle from $0\to2\pi/\kappa$:
\begin{align}
\label{eq:ca}
\delta_{\xi^-_m+\Lambda}\Q_n=\left[\Q_m,\Q_n\right]=(m-n)\Q_n+\frac{c}{12}m\left(m^2-1\right)\delta_{m+n,0},
\end{align}
from the above we see that the quantum symmetry generators form a centrally extended Virasoro algebra with regulated/normalized zero-mode $\Q_0=\frac{A}{16\pi G}$.
%--
\subsection{$AdS_2/CFT_1$ and Entropy of Near Extremal Kerr}\label{sec:ent}
%--
By employing the finite mass gauge, we have now shown that the near-extremal Kerr throat is holographically dual to a $CFT$ with center
\begin{align}\label{eq:cre}
c&=\frac{3A}{2\pi G}
\end{align}
and lowest Virasoro eigen-mode 
\begin{align}
\Q_0&=\frac{A}{16\pi G}.
\end{align}
We are now free to use the above results within the statistical Cardy Formula \eqr{eq:cf}:
\begin{align}
S=2\pi\sqrt{\frac{c\Q_0}{6}}=\frac{A}{4G}=2\pi\left(GM^2+\sqrt{G^2M^4-(GJ)^2}\right),
\end{align}
which is in agreement with the area law \eqr{eq:mal}, however derived without mixing results computed separately at near-extremality and extremality. Also the $c$ and $\Q_0$ extend smoothly to extremality in the limit as $a\to GM$ yielding:
\begin{align}
\lim_{a\to GM}c=12J~\text{and}~\lim_{a\to GM}\Q_0=J/2
\end{align}
which is the same value of the left central charge obtained in the Kerr/$CFT$ correspondence \cite{kerrcft} and together in the statistical Cardy formula the above values reproduce the extremal Kerr entropy. In addition, our derived zero-mode in \eqr{eq:cre} is in accordance with \cite{Button:2010kg} and relates to the irreducible mass of its black hole dual via:
\begin{assertion}\label{ass:zerom}
The lowest Virasoro eigen-mode of a quantum $CFT$ is proportional to the irreducible mass of its black hole dual via the form
\begin{align}\label{eq:irrmq0}
\mathcal{Q}_0=GM_{irr}^2,
\end{align}
\end{assertion}
This may be a general statement with a large avenue of application, though we are not aware of a rigorous proof at this time. 
%--
\subsection{Near Extremal Black Hole Temperature}\label{sec:temp}
%--
To extract the temperature of the $gnNHEK$ horizon, we will focus on the gravitational part of \eqr{eq:nhlcft}, i.e.
\begin{align}\label{eq:nhlcftgrav}
S_{grav}=&\frac{(r_+^2+a^2)}{16\pi G\Xi}\int d^2x\sqrt{-g^{(2)}}\left\{-\Phi\square_{g^{(2)}}\Phi+2\Phi R^{(2)}\right\}.
\end{align}
%\begin{align}
%\label{eq:louiacttem}
%S_{Liouville}=\frac{1}{96\pi}\int d^2x\sqrt{-g^{(2)}}\left\{-\Phi\square_{g^{(2)}}\Phi+2\Phi R^{(2)}\right\}.
%\end{align}
The energy momentum is given by:
\begin{align}
\label{eq:emtgra}
\bs
\left\langle T_{\mu\nu}\right\rangle=&\frac{2}{\sqrt{-g^{(2)}}}\frac{\delta S_{NHCFT}}{\delta g\indices{^{(2)}^\mu^\nu}}\\
=&\frac{r_+^2+a^2}{8\pi G}\left\{\partial_\mu\Phi\partial_\nu\Phi-2\nabla_\mu\partial_\nu\Phi+g\indices{^{(2)}_\mu_\nu}\left[2R^{(2)}-\frac12\nabla_\alpha\Phi\nabla^\alpha\Phi\right]\right\}
\es
\end{align}
and following the steps \eqr{eq:emt} through \eqr{eq:ubc}, however focusing on the horizon limit $r\to r_+$, we are left with one holomorphic component given by:
\begin{align}
\label{eq:hhf}
\left\langle T_{++}\right\rangle=-\frac{r_+^2+a^2}{32 \pi G }f'\left(r^+\right)^2.
\end{align}
This is precisely the value of the Hawking Flux of the $gnNHEK$ metric weighted by the central charge \eqr{eq:center}: 
\begin{align}
\label{eq:htfhf}
\left\langle T_{++}\right\rangle=cHF=-c\frac{\pi}{12}\left(T_H\right)^2\Rightarrow,
\end{align}
with Hawking temperature\cite{Jinwu,caldarelli:1999xj}:
\begin{align}
T_H=\frac{f'\left(r^+\right)}{4 \pi }.
\end{align}
This is an interesting result, which hints that the $AdS_2/CFT_1$ correspondence constructed here contains information about both black hole entropy and temperature. Though the $\left\langle T_{++}\right\rangle$ component in the horizon limit is not precisely the Hawking Flux of the four dimensional parent black hole, but given prior knowledge of the central extension, it is possible to read of or extract the relevant information from the correspondence.
%------------------------------------------------------------------------------
\section{$gnNHEK$ and $AdS_2$ Quantum Gravity}\label{sec:2da}
%------------------------------------------------------------------------------
We will now turn our attention to the extremal case $a=GM$ in \eqr{eq:knadsnhmkerr} for which the surface gravity vanishes identically. In other words, in contrast to the previous section we are interested in a formalism allowing $a=GM$ from the outset of the calculation, which is where the finite mass gauge will come in handy since in this limit the line element \eqr{eq:knadsnhmkerr} still exhibits an interesting Boyer-Lindquist structure. This specific case is cumbersome, as the Charge regulators of the previous section depended on $\kappa$ thus we will follow the seminal work of \cite{Castro:2009jf,Hartman:2008dq,Castro:2008ms,Chen:2011gz,Sen:2008vm,Skenderis:2002wp}. The convenient decomposition of \eqr{eq:knadsnhmkerr} into two dimensional fields should allow for an off-shell analysis of the Einstein-Hilbert action and integrating out angular degrees of freedom should leave us with an alternate effective two dimensional theory. As already seen in the previous section, the resulting effective theory should hold computational significance relevant to the near horizon regime of the near-extremal and extremal Kerr black hole.
%--
\subsection{Bulk Action}\label{sec:2dBa}
%--
Writing \eqr{eq:knadsnhmkerr} in terms of two dimensional fields and making the field redefinition $e^{-2\varphi(r)}\to e^{-2\varphi(r)}(r_+^2+a^2)$ we have
\begin{align}
\label{eq:nhnek}
ds^2=\frac{1+\cos^2\theta}{2}\left[\g{^{(2)}_\mu_\nu}dx^\mu dx^\nu+e^{-2\varphi}(r_+^2+a^2)d\theta^2\right]+\frac{2\sin^2\theta}{1+\cos^2\theta}e^{-2\varphi}(r_+^2+a^2)\left[d\phi+\mathcal{A}\right]^2
\end{align}
and substituting into the Einstein-Hilbert action
\begin{align}
\label{}
S_{EH}=\frac{1}{16\pi G}\int d^4x\sqrt{-g}R
\end{align}
and integrating out angular degrees of freedom we obtain
\begin{align}
\label{eq:ehres}
\bs
S=&\frac{2\pi}{16\pi G}\int d^2x\left\{2 e^{-4 \varphi(r)}(r_+^2+a^2)^2 \A_t'(r)^2\right.\\
&\left.-2 e^{-2\varphi(r)}(r_+^2+a^2) \left(f''(r)-4f'(r)\varphi'(r)+f(r) \left(6\varphi'(r)^2-4 \varphi''(r)\right)\right)+2\right\},
\es
\end{align}
which may be recast into covariant form
\begin{align}
\label{eq:nhagr}
\bs
S=\frac{(r_+^2+a^2)}{4G}\int d^2x\sqrt{-g^{(2)}}\left\{e^{-2\varphi}R^{(2)}+\frac{1}{(r_+^2+a^2)}+2\nabla_\mu e^{-\varphi}\nabla^\mu e^{-\varphi}-\frac{(r_+^2+a^2)}{2}e^{-4\varphi}\mc{F}^2\right\}\\
+\int d^2x\left\{\text{Total Derivative Terms}\cdots\right\}.
\es
\end{align}
Setting $\dr^2=(r_+^2+a^2)$ and dropping total derivatives we obtain the $AdS_2$ action:
\begin{align}
\label{eq:nhagder}
\bs
S_{AdS_2}=\frac{\dr^2}{4G}\int d^2x\sqrt{-g^{(2)}}\left\{e^{-2\varphi}R^{(2)}+\frac{1}{\dr^2}+2\nabla_\mu e^{-\varphi}\nabla^\mu e^{-\varphi}-\frac{\dr^2}{2}e^{-4\varphi}\mc{F}^2\right\}.
\es
\end{align}
This theory is a classical effective gravity theory and exhibits regularization via a suitable choice of boundary counterterms. This process is equivalent to renormalizing the theory on the $CFT$ side to ensure finite energy momentum and current \cite{Skenderis:2002wp}. \eqr{eq:nhagder} exhibits three equations of motion obtained by variation with respect to $\g{^{(2)}_{\mu\nu}}$, $\A$ and $\varphi.$ Remembering that for any two dimensional Riemannian metric $R_{\mu\nu}-\frac12g_{\mu\nu}R=0$ we obtain:
\begin{align}\label{eq:daseqm1}
T_{\mu\nu}=&0 &Einstein,\\\label{eq:daseqm2}
\nabla_{\mu}\mc{F}^{\mu\nu}=&0 &Maxwell,\\\label{eq:daseqm3}
R^{(2)}-e^{-2\varphi}\dr^2\mc{F}^2=&0 &constant~Scalar.
\end{align}
We will be interested in constructing the boundary counterterms to our above derived $AdS_2$ action by considering general solutions to \eqr{eq:daseqm1}-\eqr{eq:daseqm3}. A general set, as written in Gauss normal form, is given by
\begin{align}
\label{eq:gngsln}
ds^2_{(2)}=e^{-2\varphi}d\rho^2+\gamma_{tt} dt^2~\text{and}~\A=\A_\rho d\rho+\A_t dt,
\end{align}
where 
\begin{align}
\label{eq:gngslneach}
\bs
\gamma_{tt}=&-\frac14e^{-2\varphi}\left(e^{\rho/\dr}-h(t)e^{-\rho/\dr}\right)^2,\\
\A_t=&\frac{1}{2\dr}e^{\rho/\dr}\left(1-\sqrt{h(t)}e^{-\rho/\dr}\right)^2,\\
\A_\rho=&0,
\es
\end{align}
and includes a free function $h(t)$.  The $KLBH$ solution is recovered for the choice  
\begin{align}
\label{eq:hchoice}
h(t)=\frac{G^2M^2-a^2}{\dr^2}
\end{align}
and coordinate redefinition:
\begin{align}
\label{eq:coordef}
(r-GM)=\frac{\dr}{2}e^{\rho/\dr}\left(1+\frac{G^2M^2-a^2}{\dr^2}e^{-2\rho/\dr}\right).
\end{align}
The extremal case corresponds to the case $h(t)=0$.
%--
\subsection{Boundary Counterterms}\label{sec:2dBact}
%--
Following \cite{Castro:2009jf,Hartman:2008dq,Castro:2008ms}, we now determine the boundary counterterms within a well defined variational principle for \eqr{eq:nhagder}. The boundary contribution has the local form:
\begin{align}
\label{eq:Sinct}
S_{ct}=\frac{\dr^2}{2G}\int dt\sqrt{-\gamma}\left\{e^{-2\varphi}K+\alpha e^{-\varphi}+\beta e^{-3\varphi}\A_a\A^a\right\},
\end{align}
where the first term above is just the standard Gibbons-Hawking term for extrinsic curvature $K=\frac12\gamma^{tt}\sqrt{g_{\rho\rho}}\partial_\rho \gamma_{tt}=e^\varphi/\dr$, and $\alpha$ and $\beta$ are yet to be determined constants. Considering full variations including boundary terms of \eqr{eq:nhagder} we are left with solving the constraint equations:
\begin{align}
\label{eq:coneq1}
\bs
\pi_{ab}\delta\gamma^{ab}=&0,\\
\pi^{a}\delta\A_{a}=&0,\\
\pi_{\varphi}\delta\varphi=&0,
\es
\end{align}
i.e. we require vanishing canonical momenta on the boundary. Expanding the momenta on the asymptotic $AdS_2$ boundary defined by the zeroth order fields:
\begin{align}
\label{eq:ads2bcl}
\bs
\gamma\indices{^{(0)}_{tt}}=&-\frac14e^{-2\varphi^{(0)}}e^{2\rho/\dr},\\
\A\indices{^{(0)}_t}=&\frac{1}{2\dr}e^{\rho/\dr},\\
\varphi^{(0)}=&constant,
\es
\end{align}
and solving the constraint equations \eqr{eq:coneq1} we find:
\begin{align}
\label{eq:absol}
\alpha=-\frac{1}{2\dr}~\text{and}~\beta=\frac{\dr}{2}.
\end{align}
Substituting these values back into \eqr{eq:Sinct} and summarizing we obtain the total renormalized action:
\begin{align}
\label{eq:fads2act}
\bs
S_{AdS_2}^{ct}=&\frac{\dr^2}{4G}\int d^2x\sqrt{-g^{(2)}}\left\{e^{-2\varphi}R^{(2)}+\frac{1}{\dr^2}+2\nabla_\mu e^{-\varphi}\nabla^\mu e^{-\varphi}-\frac{\dr^2}{2}e^{-4\varphi}\mc{F}^2\right\}+\\
&\frac{\dr^2}{2G}\int dt\sqrt{-\gamma}\left\{e^{-2\varphi}K-\frac{1}{2\dr} e^{-\varphi}+\frac{\dr}{2} e^{-3\varphi}\A_a\A^a\right\}.
\es
\end{align}
The above action is nearly identical to the one derived in \cite{Castro:2009jf}, but differs in the coupling $\frac{\dr^2}{4G}$. 
%--
\subsection{Boundary Currents, Asymptotic Symmetries and Central Extension}\label{sec:bemtccc}
%--
The boundary energy momentum tensor and $U(1)$ current are defined as 
\begin{align}
\label{eq:emtsol}
\bs
T_{tt}=&\frac{2}{\sqrt{-\gamma}}\frac{\delta S_{AdS_2}^{ct}}{\delta \gamma\indices{^t^t}}=-\frac{\dr^2}{4G}\left(\frac{e^{-\varphi}}{\dr}\gamma_{tt}+\dr e^{-3\varphi}\A_t\A_t\right)\\
J_t=&\frac{1}{\sqrt{-\gamma}}\frac{\delta S_{AdS_2}^{ct}}{\delta \mathcal{A}_t}=\frac{\dr^2e^{-3\varphi}}{2G}\left(-e^{-\varphi}\dr^2n^\mu\mc{F}_{\mu t}+\dr\A_t\right),
\es
\end{align}
where $n^\mu$ is the radial ($\rho$) normal. We are interested in total symmetries on the $AdS_2$ boundary \eqr{eq:ads2bcl} preserving the conditions:
\begin{align}
\label{eq:gmbc}
\delta g_{\rho\rho}=\delta g_{t\rho}=0,~\delta g_{tt}=0\cdot^{2\rho/\dr},~\delta\A=0,
\end{align}
and gauge $\A_\rho=0$. A general set of diffeomorphisms preserving the above is given by:
\begin{align}
\label{eq:gmbcdiff}
\bs
\epsilon=&\left[\xi(t)+2\dr^2\left(e^{2\rho/\dr}-h(t)\right)^{-1}\xi''(t)\right]\partial_t-\dr\xi'(t)\partial_\rho\\
\Lambda=&-2\dr e^{-\rho/\dr}\left(1+\frac{G^2M^2-a^2}{\dr^2}e^{-\rho/\dr}\right)^{-1}\xi''(t),
\es
\end{align}
where $\xi(t)$ is an undetermined function of time. Expanding the boundary energy momentum tensor of \eqr{eq:emtsol} in terms of boundary fields \eqr{eq:ads2bcl} and computing its response to a total symmetry we find\footnote{The factor $\frac{1}{12}$ in the anomaly is dependent upon the choice of conformal coordinates and normalization of $\Q$. For tortoise light cone coordinates and unit normalization, as in Section~\ref{sec:qftnhnek}, the the factor is $\frac{1}{24\pi}$ \cite{cft,Iso:2008sq}.}:
\begin{align}\label{eq:bemtrp}
\bs
\delta_{\epsilon+\Lambda}T_{tt}=&2T_{tt}\xi'(t)+\xi(t)T'_{tt}+\frac{c}{12}\dr\xi'''(t)+\sqrt{h}\cdot\mc{O}\left(e^{\rho/\dr}\right),~\text{where}\\
c=&\frac{6\dr^2}{G},
\es
\end{align}
which is precisely the transformation law of an energy momentum tensor of a one dimensional  $CFT$\footnote{In contrast to \eqr{eq:clemt} and different boundary dependence.}. we should note that the factor of $\dr$ is separated in the anomalous term to ensure proper units of the central charge given by:
\begin{align}
c=\frac{6\left(r_+^2+a^2\right)}{G}=\frac{3A}{2\pi G},
\end{align}
in accord with \eqr{eq:center}. Implementing Assertion~\ref{ass:zerom} we have 
\begin{align}
\bs
c=&\frac{3A}{2\pi G},\\
\Q_0=&\frac{A}{16\pi G},
\es
\end{align}
which together inside \eqr{eq:cf} gives
\begin{align}
S=\frac{A}{4G}
\end{align}
reproducing the standard area law.

In addition, we have avoided any regulators depending on factors of $\sim1/\kappa$ or $\sim1/h$ and thus, the above analysis may be repeated for the specific case $a=GM$ from the outset, for which \eqr{eq:bemtrp} becomes:
\begin{align}%\label{eq:bemtrp}
\bs
\delta_{\epsilon+\Lambda}T_{tt}=&2T_{tt}\xi'(t)+\xi(t)T'_{tt}+\frac{c}{12}\dr\xi'''(t),~\text{where}\\
c=&12J.
\es
\end{align}
The above result is precisely the left central charge of the Kerr/$CFT$ correspondence \cite{kerrcft} obtained from the $gnNHEK$ solution in the limit $a=GM.$
%--------------------------------------------------------------------------
\section{Conclusion and Comments}\label{sec:concom}
%------------------------------------------------------------------------------
To conclude, we have analyzed quantum near-extremal Kerr black hole properties in the near horizon regime via the construction of an $AdS_2/CFT_1$ correspondence of the $gnNHEK$ metric, as outlined in Table~\ref{tb:adscftc}, and extending our previous analysis \cite{ry,Button:2010kg} to a new spacetime. The main results of our work includes the central charge $c=\frac{3A}{2\pi G}$, which was computed via a Lagrangian analysis of conserved currents of two different near horizon theories.
\begin{table}[htbp]
\begin{center}
\begin{tabular}{| c | c |}
\hline
%\multicolumn{5}{|c|}{Black Hole Comparison}\\\hline\hline
$CFT$&Black Hole \\ \hline
Conformal Group&Asymptotic Symmetry Group \\ \hline
center&$\frac{3A}{2\pi G}$ \\ \hline
Hamiltonian eigen-value&$GM^2_{irr}$ \\ \hline
Regulator&$\kappa_{gnNHEK}$ \\ \hline
\end{tabular}
\caption{Black-Hole/Near-Horizon-$CFT$ Duality}\label{tb:adscftc}
\end{center}
\end{table}

It is conceivable that other $AdS_2\times S^2$ gauges, exhibiting the field splitting of \eqr{eq:knadsnhm}, exist with relevance and physical connections to other classical near-extremal solutions. Many analogues to the $NHEK$ for charged rotating black holes with negative and positive cosmological consents have been shown to exist, see \cite{Compere:2012jk} for a comprehensive review, which suggests similar such analogues to the $gnNHEK$ solution and similar analysis of this note should be applicable. In particular, Assertion~\ref{ass:zerom} may be a useful tool in the asymptotic symmetry analysis of other extremal black holes, which have zero surface gravity and hence the need for a thermal Cardy formula \eqr{eq:tcf}. However extremal black holes in general have well defined horizons, which leads to finite non zero irreducible mass, thus allowing the implementation of a standard statistical Cardy formula \eqr{eq:cf} thus leading to the Bekenstein-Hawking entropy.
%------------------------------------------------------------------------------
\section*{Acknowledgement}
We thank Vincent Rodgers, Jacob Willig-Onwuachi, John Baker, Shanshan Rodriguez and the University of Iowa's Diffeomorphisms and Geometry research group for enlightening discussions. L.R. is grateful to the University of Iowa and NASA Goddard Space Flight Center for their hospitality during the initial and final stages of this work. 

This work is supported in part by the HHMI Undergraduate Science Education Award 52006298 and the Grinnell College Academic Affairs' CSFS and MAP programs.

%\headheight=16pt
%\rhead{Rodriguez, \textbf{\textit{D\&G}}}
%\lhead{\the\month/\the\day/\the\year }
%\lhead{L. Rodriguez}
%\chead{Notes on $M\cap\Phi$}
%\setlength{\unitlength}{1mm}
%\begin{fmffile}{fmftempl}
%------------------------------------------------------------------------
%\tableofcontents 
%\begin{center}
%\noindent\line(1,0){150}
%\end{center}

%----------------------------------------
\vspace{.5cm}
\begin{center}
\noindent\line(1,0){150}
\end{center}
%---------------------------------
\bibliographystyle{utphys}
\bibliography{cftgr}

\providecommand{\href}[2]{#2}\begingroup\raggedright\begin{thebibliography}{10}

\bibitem{hawk2}
S.~W. Hawking, ``{Particle Creation by Black Holes},''
\href{http://dx.doi.org/10.1007/BF02345020}{{\em Commun. Math. Phys.}
  {\bfseries 43} (1975) 199--220}.
%%CITATION = CMPHA,43,199;%%.

\bibitem{hawk3}
J.~M. Bardeen, B.~Carter, and S.~W. Hawking, ``{The Four laws of black hole
  mechanics},''
\href{http://dx.doi.org/10.1007/BF01645742}{{\em Commun. Math. Phys.}
  {\bfseries 31} (1973) 161--170}.
%%CITATION = CMPHA,31,161;%%.

\bibitem{beken}
J.~D. Bekenstein, ``{Black holes and entropy},''
\href{http://dx.doi.org/10.1103/PhysRevD.7.2333}{{\em Phys. Rev.} {\bfseries
  D7} (1973) 2333--2346}.
%%CITATION = PHRVA,D7,2333;%%.

\bibitem{mukwipf}
V.~F. Mukhanov, A.~Wipf, and A.~Zelnikov, ``{On 4-D Hawking radiation from
  effective action},''
  \href{http://dx.doi.org/10.1016/0370-2693(94)91255-6}{{\em Phys. Lett.}
  {\bfseries B332} (1994) 283--291},
\href{http://arxiv.org/abs/hep-th/9403018}{{\ttfamily arXiv:hep-th/9403018}}.
%%CITATION = HEP-TH/9403018;%%.

\bibitem{balfab2}
R.~Balbinot and A.~Fabbri, ``{4D quantum black hole physics from 2D models?},''
  \href{http://dx.doi.org/10.1016/S0370-2693(99)00687-5}{{\em Phys. Lett.}
  {\bfseries B459} (1999) 112--118},
\href{http://arxiv.org/abs/gr-qc/9904034}{{\ttfamily arXiv:gr-qc/9904034}}.
%%CITATION = GR-QC/9904034;%%.

\bibitem{balfab}
R.~Balbinot and A.~Fabbri, ``{Hawking radiation by effective two-dimensional
  theories},'' \href{http://dx.doi.org/10.1103/PhysRevD.59.044031}{{\em Phys.
  Rev.} {\bfseries D59} (1999) 044031},
\href{http://arxiv.org/abs/hep-th/9807123}{{\ttfamily arXiv:hep-th/9807123}}.
%%CITATION = HEP-TH/9807123;%%.

\bibitem{cadtr}
M.~Cadoni, ``{Trace anomaly and Hawking effect in generic 2D dilaton gravity
  theories},'' \href{http://dx.doi.org/10.1103/PhysRevD.53.4413}{{\em Phys.
  Rev.} {\bfseries D53} (1996) 4413--4420},
\href{http://arxiv.org/abs/gr-qc/9510012}{{\ttfamily arXiv:gr-qc/9510012}}.
%%CITATION = GR-QC/9510012;%%.

\bibitem{qpz}
S.-Q. Wu, J.-J. Peng, and Z.-Y. Zhao, ``{Anomalies, effective action and
  Hawking temperatures of a Schwarzschild black hole in the isotropic
  coordinates},'' \href{http://dx.doi.org/10.1088/0264-9381/25/13/135001}{{\em
  Class. Quant. Grav.} {\bfseries 25} (2008) 135001},
\href{http://arxiv.org/abs/0803.1338}{{\ttfamily arXiv:0803.1338 [hep-th]}}.
%%CITATION = 0803.1338;%%.

\bibitem{Camblong:2004ec}
H.~E. Camblong and C.~R. Ordonez, ``{Semiclassical methods in curved spacetime
  and black hole thermodynamics},''
  \href{http://dx.doi.org/10.1103/PhysRevD.71.124040}{{\em Phys.Rev.}
  {\bfseries D71} (2005) 124040},
\href{http://arxiv.org/abs/hep-th/0412309}{{\ttfamily arXiv:hep-th/0412309
  [hep-th]}}.
%%CITATION = HEP-TH/0412309;%%.

\bibitem{Camblong:2004ye}
H.~E. Camblong and C.~R. Ordonez, ``{Black hole thermodynamics from
  near-horizon conformal quantum mechanics},''
  \href{http://dx.doi.org/10.1103/PhysRevD.71.104029}{{\em Phys.Rev.}
  {\bfseries D71} (2005) 104029},
\href{http://arxiv.org/abs/hep-th/0411008}{{\ttfamily arXiv:hep-th/0411008
  [hep-th]}}.
%%CITATION = HEP-TH/0411008;%%.

\bibitem{Yuan:2011gq}
F.-F. Yuan and Y.-C. Huang, ``{Thermodynamics of nonspherical black holes from
  Liouville theory},''
\href{http://arxiv.org/abs/1107.5738}{{\ttfamily arXiv:1107.5738 [hep-th]}}.
%%CITATION = ARXIV:1107.5738;%%.

\bibitem{LLRphd}
L.~Rodriguez, {\em
  \href{http://ir.uiowa.edu/etd/1172}{Black-hole/near-horizon-CFT duality and 4
  dimensional classical spacetimes.}}
\newblock {Ph.D. dissertation}, University of Iowa, 2011.

\bibitem{robwill}
S.~P. Robinson and F.~Wilczek, ``{A relationship between Hawking radiation and
  gravitational anomalies},''
  \href{http://dx.doi.org/10.1103/PhysRevLett.95.011303}{{\em Phys. Rev. Lett.}
  {\bfseries 95} (2005) 011303},
\href{http://arxiv.org/abs/gr-qc/0502074}{{\ttfamily arXiv:gr-qc/0502074}}.
%%CITATION = GR-QC/0502074;%%.

\bibitem{isowill}
S.~Iso, H.~Umetsu, and F.~Wilczek, ``{Anomalies, Hawking radiations and
  regularity in rotating black holes},''
  \href{http://dx.doi.org/10.1103/PhysRevD.74.044017}{{\em Phys. Rev.}
  {\bfseries D74} (2006) 044017},
\href{http://arxiv.org/abs/hep-th/0606018}{{\ttfamily arXiv:hep-th/0606018}}.
%%CITATION = HEP-TH/0606018;%%.

\bibitem{msoda}
K.~Murata and J.~Soda, ``{Hawking radiation from rotating black holes and
  gravitational anomalies},''
  \href{http://dx.doi.org/10.1103/PhysRevD.74.044018}{{\em Phys. Rev.}
  {\bfseries D74} (2006) 044018},
\href{http://arxiv.org/abs/hep-th/0606069}{{\ttfamily arXiv:hep-th/0606069}}.
%%CITATION = HEP-TH/0606069;%%.

\bibitem{gango}
S.~Gangopadhyay, ``{Hawking radiation from black holes in de Sitter spaces via
  covariant anomalies},''
  \href{http://dx.doi.org/10.1007/s10714-009-0900-0}{{\em Gen. Rel. Grav.}
  {\bfseries 42} (2010) 1183--1187},
\href{http://arxiv.org/abs/0910.2079}{{\ttfamily arXiv:0910.2079 [hep-th]}}.
%%CITATION = 0910.2079;%%.

\bibitem{Jin}
Q.-Q. Jiang, ``{Hawking radiation from black holes in de Sitter spaces},''
  \href{http://dx.doi.org/10.1088/0264-9381/24/17/008}{{\em Class. Quant.
  Grav.} {\bfseries 24} (2007) 4391--4406},
\href{http://arxiv.org/abs/0705.2068}{{\ttfamily arXiv:0705.2068 [hep-th]}}.
%%CITATION = 0705.2068;%%.

\bibitem{Jinwu}
Q.-Q. Jiang and S.-Q. Wu, ``{Hawking radiation from rotating black holes in
  anti-de Sitter spaces via gauge and gravitational anomalies},''
  \href{http://dx.doi.org/10.1016/j.physletb.2007.01.058}{{\em Phys. Lett.}
  {\bfseries B647} (2007) 200--206},
\href{http://arxiv.org/abs/hep-th/0701002}{{\ttfamily arXiv:hep-th/0701002}}.
%%CITATION = HEP-TH/0701002;%%.

\bibitem{chen}
Z.~Xu and B.~Chen, ``{Hawking radiation from general Kerr-(anti)de Sitter black
  holes},'' \href{http://dx.doi.org/10.1103/PhysRevD.75.024041}{{\em Phys.
  Rev.} {\bfseries D75} (2007) 024041},
\href{http://arxiv.org/abs/hep-th/0612261}{{\ttfamily arXiv:hep-th/0612261}}.
%%CITATION = HEP-TH/0612261;%%.

\bibitem{chen2}
B.~Chen and W.~He, ``{Hawking Radiation of Black Rings from Anomalies},''
  \href{http://dx.doi.org/10.1088/0264-9381/25/13/135011}{{\em Class. Quant.
  Grav.} {\bfseries 25} (2008) 135011},
\href{http://arxiv.org/abs/0705.2984}{{\ttfamily arXiv:0705.2984 [gr-qc]}}.
%%CITATION = 0705.2984;%%.

\bibitem{pwu}
J.-J. Peng and S.-Q. Wu, ``{Covariant anomalies and Hawking radiation from
  charged rotating black strings in anti-de Sitter spacetimes},''
  \href{http://dx.doi.org/10.1016/j.physletb.2008.02.023}{{\em Phys. Lett.}
  {\bfseries B661} (2008) 300--306},
\href{http://arxiv.org/abs/0801.0185}{{\ttfamily arXiv:0801.0185 [hep-th]}}.
%%CITATION = 0801.0185;%%.

\bibitem{nampark}
S.~Nam and J.-D. Park, ``{Hawking radiation from covariant anomalies in 2+1
  dimensional black holes},''
  \href{http://dx.doi.org/10.1088/0264-9381/26/14/145015}{{\em Class. Quant.
  Grav.} {\bfseries 26} (2009) 145015},
\href{http://arxiv.org/abs/0902.0982}{{\ttfamily arXiv:0902.0982 [hep-th]}}.
%%CITATION = 0902.0982;%%.

\bibitem{setare}
M.~R. Setare, ``{Gauge and gravitational anomalies and Hawking radiation of
  rotating BTZ black holes},''
  \href{http://dx.doi.org/10.1140/epjc/s10052-006-0148-8}{{\em Eur. Phys. J.}
  {\bfseries C49} (2007) 865--868},
\href{http://arxiv.org/abs/hep-th/0608080}{{\ttfamily arXiv:hep-th/0608080}}.
%%CITATION = HEP-TH/0608080;%%.

\bibitem{petro}
E.~Papantonopoulos and P.~Skamagoulis, ``{Hawking Radiation via Gravitational
  Anomalies in Non- spherical Topologies},''
  \href{http://dx.doi.org/10.1103/PhysRevD.79.084022}{{\em Phys. Rev.}
  {\bfseries D79} (2009) 084022},
\href{http://arxiv.org/abs/0812.1759}{{\ttfamily arXiv:0812.1759 [hep-th]}}.
%%CITATION = 0812.1759;%%.

\bibitem{rabin3}
R.~Banerjee, ``{Covariant Anomalies, Horizons and Hawking Radiation},''
  \href{http://dx.doi.org/10.1142/S0218271808014175}{{\em Int. J. Mod. Phys.}
  {\bfseries D17} (2009) 2539--2542},
\href{http://arxiv.org/abs/0807.4637}{{\ttfamily arXiv:0807.4637 [hep-th]}}.
%%CITATION = 0807.4637;%%.

\bibitem{rabin}
R.~Banerjee and S.~Kulkarni, ``{Hawking Radiation and Covariant Anomalies},''
  \href{http://dx.doi.org/10.1103/PhysRevD.77.024018}{{\em Phys. Rev.}
  {\bfseries D77} (2008) 024018},
\href{http://arxiv.org/abs/0707.2449}{{\ttfamily arXiv:0707.2449 [hep-th]}}.
%%CITATION = 0707.2449;%%.

\bibitem{rabin2}
R.~Banerjee and S.~Kulkarni, ``{Hawking Radiation, Effective Actions and
  Covariant Boundary Conditions},''
  \href{http://dx.doi.org/10.1016/j.physletb.2007.11.068}{{\em Phys. Lett.}
  {\bfseries B659} (2008) 827--831},
\href{http://arxiv.org/abs/0709.3916}{{\ttfamily arXiv:0709.3916 [hep-th]}}.
%%CITATION = 0709.3916;%%.

\bibitem{rabin4}
R.~Banerjee and S.~Kulkarni, ``{Hawking Radiation, Covariant Boundary
  Conditions and Vacuum States},''
  \href{http://dx.doi.org/10.1103/PhysRevD.79.084035}{{\em Phys. Rev.}
  {\bfseries D79} (2009) 084035},
\href{http://arxiv.org/abs/0810.5683}{{\ttfamily arXiv:0810.5683 [hep-th]}}.
%%CITATION = 0810.5683;%%.

\bibitem{Banerjee:2008sn}
R.~Banerjee and B.~R. Majhi, ``{Connecting anomaly and tunneling methods for
  Hawking effect through chirality},''
  \href{http://dx.doi.org/10.1103/PhysRevD.79.064024}{{\em Phys. Rev.}
  {\bfseries D79} (2009) 064024},
\href{http://arxiv.org/abs/0812.0497}{{\ttfamily arXiv:0812.0497 [hep-th]}}.
%%CITATION = 0812.0497;%%.

\bibitem{Wu:2011im}
S.-Q. Wu and J.-J. Peng, ``{Thermodynamics and Hawking radiation of
  five-dimensional rotating charged Godel black holes},''
  \href{http://dx.doi.org/10.1103/PhysRevD.83.044028}{{\em Phys.Rev.}
  {\bfseries D83} (2011) 044028},
\href{http://arxiv.org/abs/1101.5474}{{\ttfamily arXiv:1101.5474 [hep-th]}}.
%%CITATION = ARXIV:1101.5474;%%.

\bibitem{Akhmedova:2008au}
V.~Akhmedova, T.~Pilling, A.~de~Gill, and D.~Singleton, ``{Comments on anomaly
  versus WKB/tunneling methods for calculating Unruh radiation},''
  \href{http://dx.doi.org/10.1016/j.physletb.2009.02.022}{{\em Phys.Lett.}
  {\bfseries B673} (2009) 227--231},
\href{http://arxiv.org/abs/0808.3413}{{\ttfamily arXiv:0808.3413 [hep-th]}}.
%%CITATION = ARXIV:0808.3413;%%.

\bibitem{Zampeli:2012tv}
A.~Zampeli, D.~Singleton, and E.~C. Vagenas, ``{Hawking radiation, chirality,
  and the principle of effective theory of gravity},''
  \href{http://dx.doi.org/10.1007/JHEP06(2012)097}{{\em JHEP} {\bfseries 1206}
  (2012) 097},
\href{http://arxiv.org/abs/1206.0879}{{\ttfamily arXiv:1206.0879 [gr-qc]}}.
%%CITATION = ARXIV:1206.0879;%%.

\bibitem{srv}
S.~Das, S.~P. Robinson, and E.~C. Vagenas, ``{Gravitational anomalies: a recipe
  for Hawking radiation},''
  \href{http://dx.doi.org/10.1142/S0218271808012218}{{\em Int. J. Mod. Phys.}
  {\bfseries D17} (2008) 533--539},
\href{http://arxiv.org/abs/0705.2233}{{\ttfamily arXiv:0705.2233 [hep-th]}}.
%%CITATION = 0705.2233;%%.

\bibitem{ry}
L.~Rodriguez and T.~Yildirim, ``{Entropy and Temperature From
  Black-Hole/Near-Horizon-CFT Duality},''
  \href{http://dx.doi.org/10.1088/0264-9381/27/15/155003}{{\em Class. Quant.
  Grav.} {\bfseries 27} (2010) 155003},
\href{http://arxiv.org/abs/1003.0026}{{\ttfamily arXiv:1003.0026 [hep-th]}}.
%%CITATION = 1003.0026;%%.

\bibitem{Button:2010kg}
B.~K. Button, L.~Rodriguez, C.~A. Whiting, and T.~Yildirim, ``{A Near Horizon
  CFT Dual for Kerr-Newman-$AdS$},''
  \href{http://dx.doi.org/10.1142/S0217751X11053663}{{\em Int.J.Mod.Phys.}
  {\bfseries A26} (2011) 3077--3090},
\href{http://arxiv.org/abs/1009.1661}{{\ttfamily arXiv:1009.1661 [hep-th]}}.
%%CITATION = ARXIV:1009.1661;%%.

\bibitem{Maldacena:1997re}
J.~M. Maldacena, ``{The large N limit of superconformal field theories and
  supergravity},'' \href{http://dx.doi.org/10.1023/A:1026654312961}{{\em Adv.
  Theor. Math. Phys.} {\bfseries 2} (1998) 231--252},
\href{http://arxiv.org/abs/hep-th/9711200}{{\ttfamily arXiv:hep-th/9711200}}.
%%CITATION = HEP-TH/9711200;%%.

\bibitem{strom2}
A.~Strominger, ``{Black hole entropy from near-horizon microstates},'' {\em
  JHEP} {\bfseries 02} (1998) 009,
\href{http://arxiv.org/abs/hep-th/9712251}{{\ttfamily arXiv:hep-th/9712251}}.
%%CITATION = HEP-TH/9712251;%%.

\bibitem{kerrcft}
M.~Guica, T.~Hartman, W.~Song, and A.~Strominger, ``{The Kerr/CFT
  Correspondence},'' \href{http://dx.doi.org/10.1103/PhysRevD.80.124008}{{\em
  Phys. Rev.} {\bfseries D80} (2009) 124008},
\href{http://arxiv.org/abs/0809.4266}{{\ttfamily arXiv:0809.4266 [hep-th]}}.
%%CITATION = 0809.4266;%%.

\bibitem{Carlip:2011ax}
S.~Carlip, ``{Extremal and nonextremal Kerr/CFT correspondences},''
  \href{http://dx.doi.org/10.1007/JHEP04(2011)076, 10.1007/JHEP01(2012)008,
  10.1007/JHEP04(2011)076, 10.1007/JHEP01(2012)008}{{\em JHEP} {\bfseries 1104}
  (2011) 076},
\href{http://arxiv.org/abs/1101.5136}{{\ttfamily arXiv:1101.5136 [gr-qc]}}.
%%CITATION = ARXIV:1101.5136;%%.

\bibitem{carlip}
S.~Carlip, ``{Black hole entropy from conformal field theory in any
  dimension},'' \href{http://dx.doi.org/10.1103/PhysRevLett.82.2828}{{\em Phys.
  Rev. Lett.} {\bfseries 82} (1999) 2828--2831},
\href{http://arxiv.org/abs/hep-th/9812013}{{\ttfamily arXiv:hep-th/9812013}}.
%%CITATION = HEP-TH/9812013;%%.

\bibitem{carlip3}
S.~Carlip, ``{Conformal field theory, (2+1)-dimensional gravity, and the BTZ
  black hole},'' \href{http://dx.doi.org/10.1088/0264-9381/22/12/R01}{{\em
  Class. Quant. Grav.} {\bfseries 22} (2005) R85--R124},
\href{http://arxiv.org/abs/gr-qc/0503022}{{\ttfamily arXiv:gr-qc/0503022}}.
%%CITATION = GR-QC/0503022;%%.

\bibitem{carlip2}
S.~Carlip, ``{Entropy from conformal field theory at Killing horizons},''
  \href{http://dx.doi.org/10.1088/0264-9381/16/10/322}{{\em Class. Quant.
  Grav.} {\bfseries 16} (1999) 3327--3348},
\href{http://arxiv.org/abs/gr-qc/9906126}{{\ttfamily arXiv:gr-qc/9906126}}.
%%CITATION = GR-QC/9906126;%%.

\bibitem{Park:1999tj}
M.-I. Park and J.~Ho, ``{Comments on 'Black hole entropy from conformal field
  theory in any dimension'},''
  \href{http://dx.doi.org/10.1103/PhysRevLett.83.5595}{{\em Phys.Rev.Lett.}
  {\bfseries 83} (1999) 5595},
\href{http://arxiv.org/abs/hep-th/9910158}{{\ttfamily arXiv:hep-th/9910158
  [hep-th]}}.
%%CITATION = HEP-TH/9910158;%%.

\bibitem{Park:2001zn}
M.-I. Park, ``{Hamiltonian dynamics of bounded space-time and black hole
  entropy: Canonical method},''
  \href{http://dx.doi.org/10.1016/S0550-3213(02)00292-4}{{\em Nucl.Phys.}
  {\bfseries B634} (2002) 339--369},
\href{http://arxiv.org/abs/hep-th/0111224}{{\ttfamily arXiv:hep-th/0111224
  [hep-th]}}.
%%CITATION = HEP-TH/0111224;%%.

\bibitem{kkp}
G.~Kang, J.-i. Koga, and M.-I. Park, ``{Near-horizon conformal symmetry and
  black hole entropy in any dimension},''
  \href{http://dx.doi.org/10.1103/PhysRevD.70.024005}{{\em Phys. Rev.}
  {\bfseries D70} (2004) 024005},
\href{http://arxiv.org/abs/hep-th/0402113}{{\ttfamily arXiv:hep-th/0402113}}.
%%CITATION = HEP-TH/0402113;%%.

\bibitem{cadss}
M.~Cadoni, ``{Statistical entropy of the Schwarzschild black hole},''
  \href{http://dx.doi.org/10.1142/S0217732306021165}{{\em Mod. Phys. Lett.}
  {\bfseries A21} (2006) 1879--1888},
\href{http://arxiv.org/abs/hep-th/0511103}{{\ttfamily arXiv:hep-th/0511103}}.
%%CITATION = HEP-TH/0511103;%%.

\bibitem{silva}
S.~Silva, ``{Black hole entropy and thermodynamics from symmetries},''
  \href{http://dx.doi.org/10.1088/0264-9381/19/15/306}{{\em Class. Quant.
  Grav.} {\bfseries 19} (2002) 3947--3962},
\href{http://arxiv.org/abs/hep-th/0204179}{{\ttfamily arXiv:hep-th/0204179}}.
%%CITATION = HEP-TH/0204179;%%.

\bibitem{bgk}
R.~Banerjee, S.~Gangopadhyay, and S.~Kulkarni, ``{Hawking radiation and near
  horizon universality of chiral Virasoro algebra},''
\href{http://dx.doi.org/10.1007/s10714-010-1028-y}{{\em Gen. Rel. Grav.}
  {\bfseries 42} (2010) 2865--2871}.
%%CITATION = GRGVA,42,2865;%%.

\bibitem{Barnes:2009zn}
E.~Barnes, D.~Vaman, and C.~Wu, ``{All 4-dimensional static, spherically
  symmetric, 2-charge abelian Kaluza-Klein black holes and their CFT duals},''
  \href{http://dx.doi.org/10.1088/0264-9381/27/9/095019}{{\em
  Class.Quant.Grav.} {\bfseries 27} (2010) 095019},
\href{http://arxiv.org/abs/0908.2425}{{\ttfamily arXiv:0908.2425 [hep-th]}}.
%%CITATION = ARXIV:0908.2425;%%.

\bibitem{Astefanesei:2009sh}
D.~Astefanesei and Y.~K. Srivastava, ``{CFT Duals for Attractor Horizons},''
  \href{http://dx.doi.org/10.1016/j.nuclphysb.2009.07.024}{{\em Nucl. Phys.}
  {\bfseries B822} (2009) 283--300},
\href{http://arxiv.org/abs/0902.4033}{{\ttfamily arXiv:0902.4033 [hep-th]}}.
%%CITATION = 0902.4033;%%.

\bibitem{Banados:2011sd}
M.~Banados and M.~Pino, ``{A note on the Cardy formula and black holes in 3d
  (massive) bigravity},''
\href{http://arxiv.org/abs/1112.0042}{{\ttfamily arXiv:1112.0042 [gr-qc]}}.
%%CITATION = ARXIV:1112.0042;%%.

\bibitem{Majhi:2011ws}
B.~R. Majhi and T.~Padmanabhan, ``{Noether Current, Horizon Virasoro Algebra
  and Entropy},'' {\em Phys.Rev.} {\bfseries D85} (2012) 084040,
\href{http://arxiv.org/abs/1111.1809}{{\ttfamily arXiv:1111.1809 [gr-qc]}}.
%%CITATION = ARXIV:1111.1809;%%.

\bibitem{Majhi:2012tf}
B.~R. Majhi and T.~Padmanabhan, ``{Noether current from the surface term of
  gravitational action, Virasoro algebra and horizon entropy},''
\href{http://arxiv.org/abs/1204.1422}{{\ttfamily arXiv:1204.1422 [gr-qc]}}.
%%CITATION = ARXIV:1204.1422;%%.

\bibitem{SheikhJabbaria:2011gc}
M.~Sheikh-Jabbari and H.~Yavartanoo, ``{EVH Black Holes, AdS3 Throats and
  EVH/CFT Proposal},'' \href{http://dx.doi.org/10.1007/JHEP10(2011)013}{{\em
  JHEP} {\bfseries 1110} (2011) 013},
\href{http://arxiv.org/abs/1107.5705}{{\ttfamily arXiv:1107.5705 [hep-th]}}.
%%CITATION = ARXIV:1107.5705;%%.

\bibitem{deBoer:2011zt}
J.~de~Boer, M.~Johnstone, M.~Sheikh-Jabbari, and J.~Simon, ``{Emergent IR Dual
  2d CFTs in Charged AdS5 Black Holes},''
  \href{http://dx.doi.org/10.1103/PhysRevD.85.084039}{{\em Phys.Rev.}
  {\bfseries D85} (2012) 084039},
\href{http://arxiv.org/abs/1112.4664}{{\ttfamily arXiv:1112.4664 [hep-th]}}.
%%CITATION = ARXIV:1112.4664;%%.

\bibitem{Yavartanoo2012410}
H.~Yavartanoo, ``On EVH black hole solution in heterotic string theory,''
  \href{http://dx.doi.org/10.1016/j.nuclphysb.2012.05.028}{{\em Nuclear Physics
  B} {\bfseries 863} no.~2, (2012) 410 -- 420}.
  \url{http://www.sciencedirect.com/science/article/pii/S0550321312003045}.

\bibitem{springerlink:10.1140/epjc/s10052-012-1911-7}
H.~Yavartanoo, ``EVH black hole solutions with higher derivative corrections,''
  {\em The European Physical Journal C - Particles and Fields} {\bfseries 72}
  (2012) 1--6. \url{http://dx.doi.org/10.1140/epjc/s10052-012-1911-7}.
  10.1140/epjc/s10052-012-1911-7.

\bibitem{rasmussen:2010xd}
J.~Rasmussen, ``{On the CFT duals for near-extremal black holes},''
  \href{http://dx.doi.org/10.1142/S0217732311035973}{{\em Mod.Phys.Lett.}
  {\bfseries A26} (2011) 1601--1611},
\href{http://arxiv.org/abs/1005.2255}{{\ttfamily arXiv:1005.2255 [hep-th]}}.
%%CITATION = ARXIV:1005.2255;%%.

\bibitem{rasmussen:2010sa}
J.~Rasmussen, ``{A near-NHEK/CFT correspondence},''
  \href{http://dx.doi.org/10.1142/S0217751X10051001}{{\em Int. J. Mod. Phys.}
  {\bfseries A25} (2010) 5517--5527},
\href{http://arxiv.org/abs/1004.4773}{{\ttfamily arXiv:1004.4773 [hep-th]}}.
%%CITATION = 1004.4773;%%.

\bibitem{Chen:2010yu}
C.-M. Chen, Y.-M. Huang, J.-R. Sun, M.-F. Wu, and S.-J. Zou, ``{On Holographic
  Dual of the Dyonic Reissner-Nordstrom Black Hole},''
  \href{http://dx.doi.org/10.1103/PhysRevD.82.066003}{{\em Phys.Rev.}
  {\bfseries D82} (2010) 066003},
\href{http://arxiv.org/abs/1006.4092}{{\ttfamily arXiv:1006.4092 [hep-th]}}.
%%CITATION = ARXIV:1006.4092;%%.

\bibitem{Chen:2010bh}
B.~Chen and J.~Long, ``{On Holographic description of the Kerr-Newman-AdS-dS
  black holes},'' \href{http://dx.doi.org/10.1007/JHEP08(2010)065}{{\em JHEP}
  {\bfseries 1008} (2010) 065},
\href{http://arxiv.org/abs/1006.0157}{{\ttfamily arXiv:1006.0157 [hep-th]}}.
%%CITATION = ARXIV:1006.0157;%%.

\bibitem{Li:2010ch}
R.~Li, M.-F. Li, and J.-R. Ren, ``{Entropy of Kaluza-Klein Black Hole from
  Kerr/CFT Correspondence},''
  \href{http://dx.doi.org/10.1016/j.physletb.2010.06.031}{{\em Phys.Lett.}
  {\bfseries B691} (2010) 249--253},
\href{http://arxiv.org/abs/1004.5335}{{\ttfamily arXiv:1004.5335 [hep-th]}}.
%%CITATION = ARXIV:1004.5335;%%.

\bibitem{Castro:2010fd}
A.~Castro, A.~Maloney, and A.~Strominger, ``{Hidden Conformal Symmetry of the
  Kerr Black Hole},'' \href{http://dx.doi.org/10.1103/PhysRevD.82.024008}{{\em
  Phys. Rev.} {\bfseries D82} (2010) 024008},
\href{http://arxiv.org/abs/1004.0996}{{\ttfamily arXiv:1004.0996 [hep-th]}}.
%%CITATION = 1004.0996;%%.

\bibitem{Krishnan:2010pv}
C.~Krishnan, ``{Hidden Conformal Symmetries of Five-Dimensional Black Holes},''
  \href{http://dx.doi.org/10.1007/JHEP07(2010)039}{{\em JHEP} {\bfseries 07}
  (2010) 039},
\href{http://arxiv.org/abs/1004.3537}{{\ttfamily arXiv:1004.3537 [hep-th]}}.
%%CITATION = 1004.3537;%%.

\bibitem{kerrcftstring}
T.~Azeyanagi, N.~Ogawa, and S.~Terashima, ``{The Kerr/CFT Correspondence and
  String Theory},'' \href{http://dx.doi.org/10.1103/PhysRevD.79.106009}{{\em
  Phys. Rev.} {\bfseries D79} (2009) 106009},
\href{http://arxiv.org/abs/0812.4883}{{\ttfamily arXiv:0812.4883 [hep-th]}}.
%%CITATION = 0812.4883;%%.

\bibitem{kerrcftsugra}
D.~D.~K. Chow, M.~Cvetic, H.~Lu, and C.~N. Pope, ``{Extremal Black Hole/CFT
  Correspondence in (Gauged) Supergravities},''
  \href{http://dx.doi.org/10.1103/PhysRevD.79.084018}{{\em Phys. Rev.}
  {\bfseries D79} (2009) 084018},
\href{http://arxiv.org/abs/0812.2918}{{\ttfamily arXiv:0812.2918 [hep-th]}}.
%%CITATION = 0812.2918;%%.

\bibitem{kerrcftind}
H.~Lu, J.~Mei, and C.~N. Pope, ``{Kerr/CFT Correspondence in Diverse
  Dimensions},'' \href{http://dx.doi.org/10.1088/1126-6708/2009/04/054}{{\em
  JHEP} {\bfseries 04} (2009) 054},
\href{http://arxiv.org/abs/0811.2225}{{\ttfamily arXiv:0811.2225 [hep-th]}}.
%%CITATION = 0811.2225;%%.

\bibitem{daCunha:2010jj}
B.~C. da~Cunha and A.~R. de~Queiroz, ``{Kerr-CFT From Black-Hole
  Thermodynamics},'' \href{http://dx.doi.org/10.1007/JHEP08(2010)076}{{\em
  JHEP} {\bfseries 08} (2010) 076},
\href{http://arxiv.org/abs/1006.0510}{{\ttfamily arXiv:1006.0510 [hep-th]}}.
%%CITATION = 1006.0510;%%.

\bibitem{Wu:2009di}
X.-N. Wu and Y.~Tian, ``{Extremal Isolated Horizon/CFT Correspondence},'' {\em
  Phys.Rev.} {\bfseries D80} (2009) 024014,
\href{http://arxiv.org/abs/0904.1554}{{\ttfamily arXiv:0904.1554 [hep-th]}}.
%%CITATION = ARXIV:0904.1554;%%.

\bibitem{Huang:2010yg}
Y.-C. Huang and F.-F. Yuan, ``{Hidden conformal symmetry of extremal
  Kaluza-Klein black hole in four dimensions},''
  \href{http://dx.doi.org/10.1007/JHEP03(2011)029}{{\em JHEP} {\bfseries 1103}
  (2011) 029},
\href{http://arxiv.org/abs/1012.5453}{{\ttfamily arXiv:1012.5453 [hep-th]}}.
%%CITATION = ARXIV:1012.5453;%%.

\bibitem{Compere:2012jk}
G.~Compere, ``{The Kerr/CFT correspondence and its extensions: a comprehensive
  review},''
\href{http://arxiv.org/abs/1203.3561}{{\ttfamily arXiv:1203.3561 [hep-th]}}.
%%CITATION = ARXIV:1203.3561;%%.

\bibitem{cardy2}
H.~W.~J. Bl{\"o}te, J.~A. Cardy, and M.~P. Nightingale {\em Phys. Rev. Lett.}
  {\bfseries 56} no.~742, (1986) .

\bibitem{cardy1}
J.~A. Cardy {\em Nucl. Phys.} {\bfseries B270} no.~186, (1986) .

\bibitem{frolovthorne}
V.~P. Frolov and K.~S. Thorne, ``{Renormalized Stress - Energy Tensor near the
  Horizon of a Slowly Evolving, Rotating Black Hole},''
\href{http://dx.doi.org/10.1103/PhysRevD.39.2125}{{\em Phys. Rev.} {\bfseries
  D39} (1989) 2125--2154}.
%%CITATION = PHRVA,D39,2125;%%.

\bibitem{ChangYoung:2012kd}
E.~Chang-Young and M.~Eune, ``{Nonextremal Kerr/CFT on a stretched horizon},''
\href{http://arxiv.org/abs/1212.3031}{{\ttfamily arXiv:1212.3031 [hep-th]}}.
%%CITATION = ARXIV:1212.3031;%%.

\bibitem{Castro:2009jf}
A.~Castro and F.~Larsen, ``{Near Extremal Kerr Entropy from $AdS_2$ Quantum
  Gravity},'' \href{http://dx.doi.org/10.1088/1126-6708/2009/12/037}{{\em JHEP}
  {\bfseries 12} (2009) 037},
\href{http://arxiv.org/abs/0908.1121}{{\ttfamily arXiv:0908.1121 [hep-th]}}.
%%CITATION = 0908.1121;%%.

\bibitem{Bardeen:1999px}
J.~M. Bardeen and G.~T. Horowitz, ``{The Extreme Kerr throat geometry: A Vacuum
  analog of $AdS_2\times S^2$},''
  \href{http://dx.doi.org/10.1103/PhysRevD.60.104030}{{\em Phys.Rev.}
  {\bfseries D60} (1999) 104030},
\href{http://arxiv.org/abs/hep-th/9905099}{{\ttfamily arXiv:hep-th/9905099
  [hep-th]}}.
%%CITATION = HEP-TH/9905099;%%.

\bibitem{rodrigues:2012emd}
M.~E. Rodrigues and G.~T. Marques, ``{Thermodynamics of a class of
  non-asymptotically flat black holes in Einstein-Maxwell-Dilaton theory},''
\href{http://arxiv.org/abs/1206.0763}{{\ttfamily arXiv:1206.0763 [gr-qc]}}.
%%CITATION = ARXIV:1206.0763;%%.

\bibitem{wald}
R.~Wald, {\em General Relativity}.
\newblock University of Chicago Press, 1984.

\bibitem{Amsel:2009ev}
A.~J. Amsel, G.~T. Horowitz, D.~Marolf, and M.~M. Roberts, ``{No Dynamics in
  the Extremal Kerr Throat},'' {\em JHEP} {\bfseries 0909} (2009) 044,
\href{http://arxiv.org/abs/0906.2376}{{\ttfamily arXiv:0906.2376 [hep-th]}}.
%%CITATION = ARXIV:0906.2376;%%.

\bibitem{Maldacena:1998uz}
J.~M. Maldacena, J.~Michelson, and A.~Strominger, ``{Anti-de Sitter
  fragmentation},'' {\em JHEP} {\bfseries 9902} (1999) 011,
\href{http://arxiv.org/abs/hep-th/9812073}{{\ttfamily arXiv:hep-th/9812073
  [hep-th]}}.
%%CITATION = HEP-TH/9812073;%%.

\bibitem{Bredberg:2009pv}
I.~Bredberg, T.~Hartman, W.~Song, and A.~Strominger, ``{Black Hole
  Superradiance From Kerr/CFT},''
  \href{http://dx.doi.org/10.1007/JHEP04(2010)019}{{\em JHEP} {\bfseries 1004}
  (2010) 019},
\href{http://arxiv.org/abs/0907.3477}{{\ttfamily arXiv:0907.3477 [hep-th]}}.
%%CITATION = ARXIV:0907.3477;%%.

\bibitem{2000GReGr..32.1665P}
A.~Z. {Petrov}, ``{The Classification of Spaces Defining Gravitational
  Fields},'' \href{http://dx.doi.org/10.1023/A:1001910908054}{{\em General
  Relativity and Gravitation} {\bfseries 32} (Aug., 2000) 1665--1685}.

\bibitem{Yale:2010tn}
A.~Yale, ``{Exact Hawking Radiation of Scalars, Fermions, and Bosons Using the
  Tunneling Method Without Back-Reaction},''
  \href{http://dx.doi.org/10.1016/j.physletb.2011.02.023}{{\em Phys.Lett.}
  {\bfseries B697} (2011) 398--403},
\href{http://arxiv.org/abs/1012.3165}{{\ttfamily arXiv:1012.3165 [gr-qc]}}.
%%CITATION = ARXIV:1012.3165;%%.

\bibitem{Chung:2010xy}
H.~Chung, ``{Dynamics of Diffeomorphism Degrees of Freedom at a Horizon},''
  \href{http://dx.doi.org/10.1103/PhysRevD.83.084017}{{\em Phys. Rev.}
  {\bfseries D83} (2011) 084017},
\href{http://arxiv.org/abs/1011.0623}{{\ttfamily arXiv:1011.0623 [gr-qc]}}.
%%CITATION = 1011.0623;%%.

\bibitem{Chung:2010xz}
H.~Chung, ``{Hawking Radiation and Entropy from Horizon Degrees of Freedom},''
  \href{http://dx.doi.org/10.1016/j.nuclphysb.2012.01.011}{{\em Nucl.Phys.}
  {\bfseries B858} (2012) 214--231},
\href{http://arxiv.org/abs/1011.0624}{{\ttfamily arXiv:1011.0624 [gr-qc]}}.
%%CITATION = ARXIV:1011.0624;%%.

\bibitem{solodukhin:1998tc}
S.~N. Solodukhin, ``{Conformal description of horizon's states},''
  \href{http://dx.doi.org/10.1016/S0370-2693(99)00398-6}{{\em Phys. Lett.}
  {\bfseries B454} (1999) 213--222},
\href{http://arxiv.org/abs/hep-th/9812056}{{\ttfamily arXiv:hep-th/9812056}}.
%%CITATION = HEP-TH/9812056;%%.

\bibitem{strom1}
A.~Strominger, ``{Les Houches lectures on black holes},'' 1994.

\bibitem{Leutwyler:1984nd}
H.~Leutwyler, ``{Gravitational Anomalies: A Soluble Two-Dimensional Model},''
\href{http://dx.doi.org/10.1016/0370-2693(85)91443-1}{{\em Phys. Lett.}
  {\bfseries B153} (1985) 65}.
%%CITATION = PHLTA,B153,65;%%.

\bibitem{unruh}
W.~G. Unruh, ``{Notes on black hole evaporation},''
\href{http://dx.doi.org/10.1103/PhysRevD.14.870}{{\em Phys. Rev.} {\bfseries
  D14} (1976) 870}.
%%CITATION = PHRVA,D14,870;%%.

\bibitem{cft}
P.~D. Francesco, P.~Mathieu, and D.~S{\'e}n{\'e}chal, {\em Conformal Field
  Theory}.
\newblock Springer, 1997.

\bibitem{caldarelli:1999xj}
M.~M. Caldarelli, G.~Cognola, and D.~Klemm, ``{Thermodynamics of
  Kerr-Newman-AdS black holes and conformal field theories},''
  \href{http://dx.doi.org/10.1088/0264-9381/17/2/310}{{\em Class. Quant. Grav.}
  {\bfseries 17} (2000) 399--420},
\href{http://arxiv.org/abs/hep-th/9908022}{{\ttfamily arXiv:hep-th/9908022}}.
%%CITATION = HEP-TH/9908022;%%.

\bibitem{Hartman:2008dq}
T.~Hartman and A.~Strominger, ``{Central Charge for AdS(2) Quantum Gravity},''
  \href{http://dx.doi.org/10.1088/1126-6708/2009/04/026}{{\em JHEP} {\bfseries
  0904} (2009) 026},
\href{http://arxiv.org/abs/0803.3621}{{\ttfamily arXiv:0803.3621 [hep-th]}}.
%%CITATION = ARXIV:0803.3621;%%.

\bibitem{Castro:2008ms}
A.~Castro, D.~Grumiller, F.~Larsen, and R.~McNees, ``{Holographic Description
  of AdS(2) Black Holes},''
  \href{http://dx.doi.org/10.1088/1126-6708/2008/11/052}{{\em JHEP} {\bfseries
  0811} (2008) 052},
\href{http://arxiv.org/abs/0809.4264}{{\ttfamily arXiv:0809.4264 [hep-th]}}.
%%CITATION = ARXIV:0809.4264;%%.

\bibitem{Chen:2011gz}
C.-M. Chen and J.-R. Sun, ``{Holographic Dual of the Reissner-Nordstr\"om Black
  Hole},'' {\em J.Phys.Conf.Ser.} {\bfseries 330} (2011) 012009,
\href{http://arxiv.org/abs/1106.4407}{{\ttfamily arXiv:1106.4407 [hep-th]}}.
%%CITATION = ARXIV:1106.4407;%%.

\bibitem{Sen:2008vm}
A.~Sen, ``{Quantum Entropy Function from AdS(2)/CFT(1) Correspondence},''
  \href{http://dx.doi.org/10.1142/S0217751X09045893}{{\em Int.J.Mod.Phys.}
  {\bfseries A24} (2009) 4225--4244},
\href{http://arxiv.org/abs/0809.3304}{{\ttfamily arXiv:0809.3304 [hep-th]}}.
%%CITATION = ARXIV:0809.3304;%%.

\bibitem{Skenderis:2002wp}
K.~Skenderis, ``{Lecture notes on holographic renormalization},''
  \href{http://dx.doi.org/10.1088/0264-9381/19/22/306}{{\em Class.Quant.Grav.}
  {\bfseries 19} (2002) 5849--5876},
\href{http://arxiv.org/abs/hep-th/0209067}{{\ttfamily arXiv:hep-th/0209067
  [hep-th]}}.
%%CITATION = HEP-TH/0209067;%%.

\bibitem{Iso:2008sq}
S.~Iso, ``{Hawking Radiation, Gravitational Anomaly and Conformal Symmetry: The
  Origin of Universality},''
  \href{http://dx.doi.org/10.1142/S0217751X08040627}{{\em Int.J.Mod.Phys.}
  {\bfseries A23} (2008) 2082--2090},
\href{http://arxiv.org/abs/0804.0652}{{\ttfamily arXiv:0804.0652 [hep-th]}}.
%%CITATION = ARXIV:0804.0652;%%.

\end{thebibliography}\endgroup
%\nocite{*}
%---------------------------------

\end{document}